\documentclass{aa}  

\usepackage{amsmath,amssymb}
\usepackage[breaklinks=true]{hyperref}
\newcommand{\orcit}[1]{\protect\href{https://orcid.org/#1}{\protect\includegraphics[width=8pt]{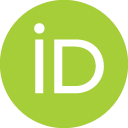}}}
\usepackage{graphicx}
\usepackage{xcolor}
\usepackage[normalem]{ulem}
\usepackage{listings}
\definecolor{dkgreen}{rgb}{0,0.6,0}
\definecolor{gray}{rgb}{0.5,0.5,0.5}
\definecolor{mauve}{rgb}{0.58,0,0.82}
\usepackage{txfonts}
\def\deg{\ensuremath{^\circ}}

\providecommand{\kms}{\ensuremath{\rm \,km\,s^{-1}}\xspace}

\providecommand{\kpc}{\ensuremath{\,\rm kpc}\xspace}

\providecommand{\kms}{\ensuremath{\textrm{km\,s}^{-1}}}

\providecommand{\degree}{\ensuremath{^\circ}}

\providecommand{\gaia}{\textit{Gaia}}

\newcommand{\drimmelgaia}{{GD23}}

\usepackage{graphicx}
\usepackage{txfonts}
\usepackage{hyperref}

\makeatletter
\renewcommand*\maketitle{%
  \thispagestyle{firstpage}
\begingroup
    \if@wideboxfn
    \setlength\bibindent{1.4\parindent}
    \else
    \setlength\bibindent{\parindent}
    \fi
    \renewcommand*\thefootnote{\@fnsymbol\c@footnote}%
    \renewcommand\@makefntext[1]{%
    \ifaa@longfn\hsize\textwidth\fi
    \noindent
    \hb@xt@\bibindent{\hss\@makefnmark\enspace}##1}
  \ifaa@twocolumn
  \begingroup
    \begin{aa@strip}
          \aa@maketitle
    \end{aa@strip}
    \@thanks            
  \endgroup
  \else
    \begingroup
      \let\thanks\footnote
      \aa@maketitle
    \endgroup
  \fi
\endgroup
  \setcounter{footnote}{0}%
}
\makeatother
\makeatletter
\DeclareRobustCommand*{\fieldName}[1]{%
  \begingroup\@fieldName\scantokens{\texttt{\small {#1}}\noexpand}\endgroup}
\begingroup\lccode`\~=`\_\relax
   \lowercase{\endgroup\def\@fieldName{\catcode`\_=\active \let~\_}}
\makeatother

\begin{document} 

   \title{The Milky Way as seen by classical Cepheids II: Spiral structure
   }

\author{ 
R. Drimmel\orcit{0000-0002-1777-5502}\inst{1}  \and
S. Khanna\orcit{0000-0002-2604-4277}\inst{1} \and
E. Poggio\orcit{0000-0003-3793-8505}\inst{1} \and
D. M. Skowron\orcit{[0000-0001-9439-604X]}\inst{2} 
}

\institute{INAF - Osservatorio Astrofisico di Torino, via Osservatorio 20, 10025 Pino Torinese (TO), Italy\\    \email{ronald.drimmel@inaf.it}
        \and
Astronomical Observatory, University of Warsaw, Al. Ujazdowskie 4, 00-478 Warsaw, Poland
         }
         
       \date{Received ; accepted }

 
  \abstract{  
As a relatively young and bright population and the archetype of standard candles, classical Cepheids are an ideal population on which to trace the non-axisymmetric structure in the young stellar disk to large distances. We used the new distances derived in Paper I based on mid-IR WISE photometry for a selected sample of 2857 dynamically young Cepheids to trace the spiral arms of the Milky Way. 
The Perseus and Sagittarius-Carina arms are clearly evident in the third and fourth Galactic quadrants, while the Local and Scutum arms are much weaker, and extinction severely limits our view of the latter innermost spiral arm. Pitch angles were derived for each arm over various ranges of Galactic azimuth, each covering at least 90\deg\ in azimuth. Our method of detecting spiral arms and deriving pitch angles does not rely on pre-assigning sources to specific arms. 
While the spiral structure in the first and second quadrant is not obvious in part because of extinction effects, it is not inconsistent with the structure seen in the third and fourth quadrants. In summary, the Cepheids allow us to map spiral structure in the third and fourth Galactic quadrants where currently few masers have astrometric parallaxes, significantly extending our understanding of the Milky Way at large scales. 
  }

 
   \maketitle
%

\section{Introduction}


Mapping the large-scale spiral structure of the Milky Way as traced by star formation has been a perpetual challenge because we are located within a dust-filled disk. In the optical, we are limited to about four to five kiloparsecs from the Sun using bright young stars as seen by \gaia\ \citep[][hereafter \drimmelgaia{}]{Xu:2018, Zari2021,Poggio2021,Drimmel23_gaia}.  The absolute astrometry of maser radio sources, which are identified as high-mass stars hosting circumstellar disks, has allowed major progress in the past decade by mapping the first and second Galactic quadrants \citep[][]{Reid2019,VERA2020}.  However, the number of masers with measured parallaxes in the third and fourth quadrants, which are mostly visible from the southern hemisphere, is still quite limited and does not provide sufficient sources to confidently map the location of the star formation complexes that cause the spiral structure that we would see in the optical. For a recent review attempting to integrate \textit{Gaia} astrometry for young stars and data from radio masers, we refer to \citet{Xu2023}.

We investigate the spiral structure as seen by classical Cepheids, which are a predominantly young population of stars whose distances can be reliably found because they are the archetype of the standard candle.  Previous studies of spiral structure using Cepheids include \citet{Griv2017}, who used 674 Cepheids within about 3kpc and identified three arm segments with pitch angles of 12, 10, and 17\deg, using a Fourier decomposition of three log-spiral components.  
\citet{Skowron2019a,Skowron2019b} showed that the distribution of 2390 Cepheids 
in the Galactic plane is consistent with the majority being born in spiral arms after diffusion effects and the radial age-gradient of the Cepheids were taken into account. However, no attempt was made to map or parametrise the spiral arms themselves.  More recently,
\citet{Minniti2021} used 50 Classical Cepheids in a first  attempt to model the spiral arms on the far side of the Galaxy. 

Even more recent attempts to map the spiral structure with the classical Cepheids of the Milky Way include 
\citet{Lemasle2022}, who used a sample of 2684 Cepheids compiled from various variability catalogues, with distances derived using a period-Wesenheit relation in the Wide-field Infrared Survey Explorer (WISE) bands and \gaia\ parallaxes for Cepheids without unWISE \citep{Schlafly2019} photometry. 
Primarily based on the subsample of Cepheids younger than 150~Myr, they identified numerous spiral arm segments.  Moreover, \citet{Bobylev2022} used about 600 pre-selected Cepheids from \citet{Skowron2019a,Skowron2019b} to measure the pitch angle of the Sgr-Car arm and an outer arm beyond Perseus.  
\drimmelgaia{} used 2808 young classical Cepheids to construct an overdensity map of their distribution, identifying the Sagittarius and Perseus arms, but making no attempt to parametrise the arms. 

The identification of the Cepheids is dependent on observations made in the optical where their variability follows a characteristic profile that is needed for identification. Our distances instead rely on $W1$ mid-infrared (MIR) photometry from the WISE satellite \citep{Wright2010}.  Relying on MIR photometry minimises the effect of extinction, which must be taken into account when deriving the photometric distances, and also the uncertainty from the intrinsic scatter about the PL relation \citep{Bono2024}.  Instead of using Wesenheit-based distances, we here use the recent Cepheid distances derived from WISE MIR $W1$ photometry that rely on the PL relation in $W1$ and an extinction model \citep[][hereafter Paper~I]{Skowron2024}. 
A comparison between these and Wesenheit-based distances shows systematic differences towards the inner disk of the Milky Way that can be attributed to unaccounted-for variations in the extinction curve. These new distances were validated on a set of 910 Cepheids with good astrometry and are shown to have relative distance uncertainties smaller than 13\%.

The paper is organised as follows: Section 2 describes our selection of young Cepheids that we used as spiral tracers, and Sect. 3 explains the method we used to identify and characterise spiral arms in the distribution of our sample.  In Sect. 4 we present the results of our analysis over various ranges of Galactic azimuth angle, and we consider alternative age-selection criteria. Finally, in Sect. 5 we discuss our results in the context of previous efforts to map the Galactic spiral structure on large scales, and we briefly summarise our results in Sect. 6.

\section{Data}
\label{sec:data}


To trace the young population, we used the sample of 3659 known classical Cepheids from \citet{Pietrukowicz:2021}, for which we derived new distances for 3425 Cepheids  based on WISE $W1$ photometry from the AllWISE \citep{Cutri2013} and unWISE \citep{Schlafly2019} catalogues in Paper~I.  For details of the sample definition and the distance derivation, we refer to Paper~I. 

Finally, we removed from this catalogue of Cepheid distances objects whose distances are clearly inconsistent with their astrometry based on the quantity provided in the catalogue,
\begin{equation}
    Q = |\Delta \varpi / \sigma_{\Delta \varpi}|\,, 
\end{equation}
where $\Delta \varpi = \varpi - \varpi_\mu$, where $\varpi_\mu$ is the photometric parallax for a source with a distance modulus $\mu$, that is, 
\begin{equation}
    \varpi_\mu = 10^{-(\mu - 10)/5}\,,
\end{equation}
and $\sigma_{\Delta \varpi}$ the estimated uncertainty of $\Delta \varpi$ (see Paper~I for details).  We imposed the condition that $Q < 5$.  This removed 63 presumable contaminants from our sample and left a total of 3362 Cepheids.


For age estimates of our Cepheid sample we used the ages provided in PaperI, which are based on the period-age-metallicity relation derived by \citet{Anderson2016}, who took the effect of stellar rotation into account (see PaperI for further details).  While these ages may be quite uncertain for individual Cepheids, we assumed that they are sufficiently accurate to select a sample.  

Since Cepheids can span a significant range of ages but we wished to use them as a tracer of a young stellar population, we selected them based on age. Common age cuts found in the literature for this purpose are typically 200 Myr or younger. However, because of the metallicity gradient, Cepheids in the outer Galaxy are systematically older than those in the inner Galaxy because low-mass stars in metal-rich environments do not reach the Cepheid instability strip during their evolution \citep{Skowron2019a,Anders:2024}.  As a result, a simple age selection excludes many Cepheids in the outer disk. In addition, such an age cut does not take into account that dynamical timescales are very different in the outer Galaxy with respect to the inner Galaxy. Cepheids as old as 200~Myr or more may still not have had time to wander far from their birth radius in the outer Galaxy, while Cepheids of the same age in the inner Galaxy may have already completed nearly two Galactic rotations. 

As an alternative to a simple age cut, we selected the objects based on the dynamical age of the Cepheids, that is, their age with respect to the epicyclic frequency at their current galactocentric radius. We thus required their age to be younger than 
the epicyclic period ($= 2\pi/\kappa$), where the epicyclic frequency $\kappa$ in the epicyclic approximation is
\begin{equation}
    \kappa^2(R_g) = \left( R \frac{d \Omega^2}{dR} + 4 \Omega^2 \right)_{R_g}\, ,
\end{equation}
where $R$ is the galactocentric cylindrical radius, and the angular velocity is $\Omega^2 = V_\phi^2/R^2$. Evaluating $\Omega^2$ at the guiding radius $R_g$, we can substitute $\Omega(R_g)^2 = L^2_Z/R_g^4$, where $L_Z$ is the vertical angular momentum component. \citet{Drimmel2023} showed that the angular momentum $L_Z$ of the subset of DR3 Cepheids with line-of-sight velocities follows a simple linear relation with respect to galactocentric radius $R$, that is, $L_Z(R) = 231.4$ \kms $\times R$, over a range of radii $5 < R < 18$ \kpc, implying a flat rotation curve of 231.4 \kms over this radial extent. Taking $L_Z$ as a proxy for $R_g$, we thus find $\Omega(R_g) = 231.4/R_g$ for the angular velocity of the Cepheids, and the epicyclic frequency is then $\kappa(R_g) = \sqrt{2} \cdot 231.4/R_g$.
Making the further assumption that the guiding radius $R_g$ can be substituted by $R$ (approximately true for a genuinely young population), our age criterion becomes
\begin{equation}
    Age_{Myr} < \frac{\sqrt{2} \pi R}{0.2314} ,
    \label{age_crit}
\end{equation}
for $R$ in \kpc and using the convenient approximation $1 \kms \simeq 1$pc per Myr.
At the solar radius, this requires Cepheids to be younger than about 160~Myr to be considered dynamically young.
A more accurate determination of $R_g$ per star would require that line-of-sight velocities are available. 

To apply the above criterion (Eq. \ref{age_crit}), we just need the galactocentric cylindrical radius $R$, which we derived using the most recent derivation of the distance to the Sun from the Galactic centre, $R_\odot$, of the ESO Gravity project, which is based on the measurements of stars orbiting the supermassive black hole Sgr A* at the Galactic centre, namely $R_\odot = 8277 \pm 9$(stat) $\pm 30$(sys) pc \citep[][]{GravityCollaboration:2022}. The heliocentric Cartesian coordinates are defined by 
\begin{equation}
\begin{pmatrix}
    x \\
    y \\
    z
\end{pmatrix} =
\begin{pmatrix}
d \cos l \cos b  \\
d \sin l \cos b \\
d \sin b
\end{pmatrix}\,,
\end{equation}
for galactic coordinates $(l,b)$ and a heliocentric distance $d$.
We translated this into galactocentric Cartesian coordinates $(X,Y,Z)$ using $R_\odot$, and for simplicity, assumed $Z_\odot = 0$. We also kept the $X$-axis as pointing in the direction of the Galactic centre as seen from the Sun. That is, $X = x - R_\odot$, $Y=y$, and $Z = z$. The galactocentric cylindrical radius $R$ is then $\sqrt{X^2+Y^2}$, and the galactocentric azimuth was taken as $\phi={\rm tan}^{-1}(Y/X)$. The galactocentric azimuth defined in this way increases in the anti-clockwise direction, while the rotation of the Galaxy is clockwise, as seen from the north Galactic pole.

\begin{figure}
    \centering   \includegraphics[width=0.49\textwidth]{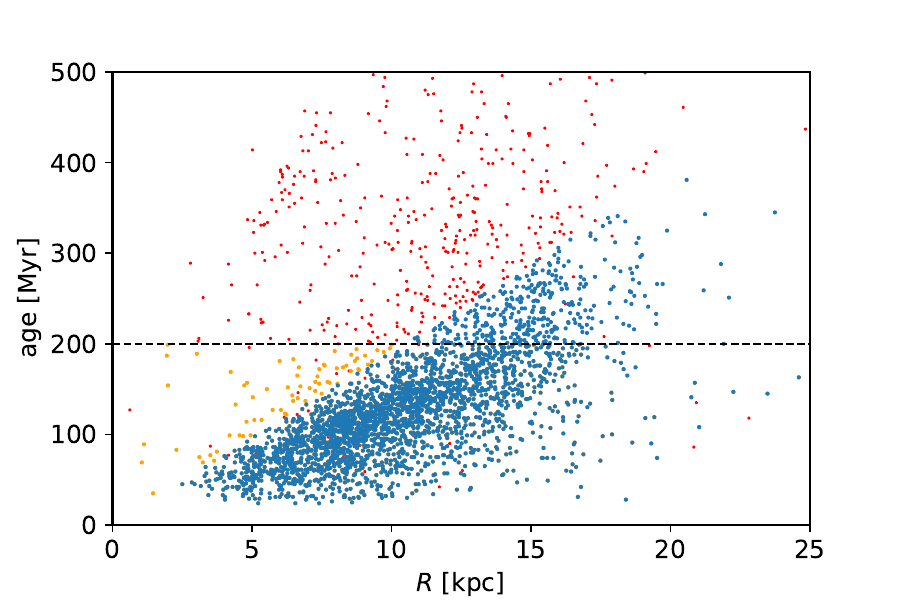}
    \caption{Distribution of Cepheid ages with respect to galactocentric radius. The blue points show dynamically young Cepheids, yellow points show Cepheids that are younger than 200~Myr but not dynamically young, and red points are Cepheids older than 200~Myr that are not dynamically young either.
    }
    \label{fig:ages}
\end{figure}

Figure~\ref{fig:ages} shows the age of the Cepheids with respect to galactocentric radius and our age selection. The dynamically young criterion becomes a more stringent cut at smaller galactocentric radii, but the mean age gradient of Cepheids mentioned above means that most Cepheids are dynamically young in any case. In contrast, in the outer Galaxy, our new age criterion allows us to include many  Cepheids that would otherwise be excluded from our sample with a simple age criterion. Of our 2857 dynamically young Cepheids, 374 are older than 200~Myr.  

\section{Identification and characterisation of spiral arms}
\label{sec:spirals}

\begin{figure}
    \centering
    \includegraphics[width=0.49\textwidth]{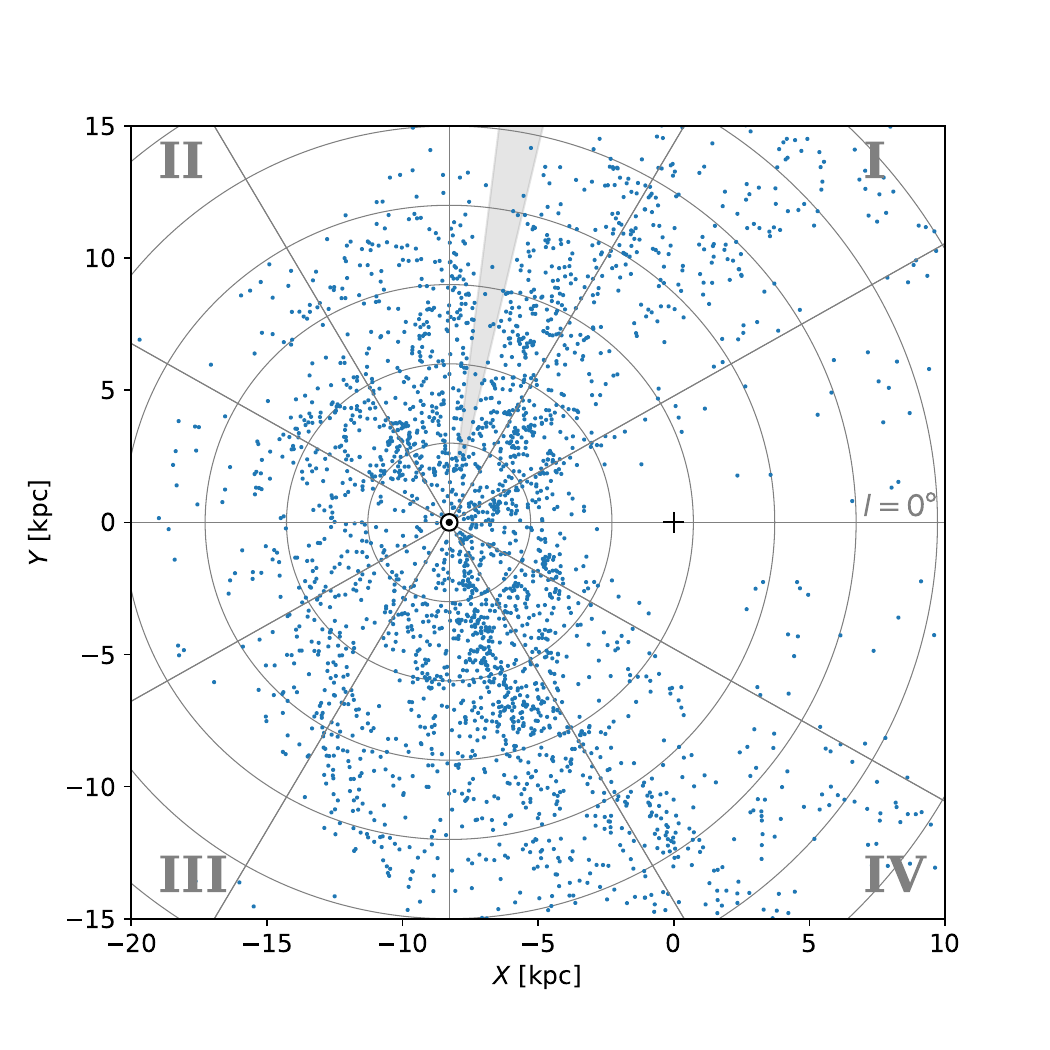}
    \caption{Distribution of 2857 dynamically young Cepheids in the Galactic plane. The dark blue points show the positions of the dynamically young Cepheids.
    The positions of the Galactic centre (+) and the Sun ($\odot$) are indicated, and the overlaid polar grid shows the galactic longitude in 30\deg\ intervals and heliocentric distances at 3\kpc intervals. The shaded region shows the shadow cone caused by high extinction in the Cygnus X star formation region. The Galactic quadrants are indicated with Roman numerals.
    }
    \label{fig:XY}
\end{figure}

We now use the dynamically young Cepheids to map the spiral arms. Figure~\ref{fig:XY} shows the positions of the dynamically young Cepheids in the Galactic plane in galactocentric coordinates, which shows a spiral arm that is clearly traced by the Cepheids in the fourth quadrant ($270\deg < l < 360\deg$), just inside the position of the Sun. As this arm can be clearly traced from $l=0\deg$ (Sagittarius) to the arm tangent in the direction of Carina ($l \approx 285\deg$), we identify this arm as the Sagittarius-Carina (Sgr-Car) arm.  Another weaker and broader arm is also dimly visible in the third quadrant ($180\deg < l < 270\deg$). In the first and second quadrants, however, the distribution of Cepheids shows no obvious spiral structure, but is much more discontinuous and patchy.  The lack of obvious spiral features in the first and second quadrants is in part due to the non-uniform coverage in this half of the Galaxy due to interstellar extinction, which introduces gaps or shadow cones along lines of sight with strong foreground extinction. Some of these shadow cones are easily recognised in Fig.~\ref{fig:XY}, in particular, one cone in the direction of $l \approx 80\deg$ that starts at about 2-3 \kpc from the Sun, where the Cygnus X star-forming region is located (shown as a shaded region in Fig.~\ref{fig:XY}).  Extinction has a greater affect in the first and second quadrants in part because of the geometry of the spiral arms: In these directions, our lines of sight cross spiral arms in a closer vicinity than in the third and fourth quadrant, where the separation of the arms is larger. This allows us an unhindered view over wider ranges in galactic longitude.  The dust lanes of a spiral arm can limit the view of the arm itself and of any other arms beyond it.  However, not all patchiness can be explained by extinction. It also arises to some extent from star formation, which itself is patchy, and does not occur continuously along spiral arms. Finally, we recall that our sample is based on a set of inhomogeneous catalogues and therefore has a selection function that is the sum result of overlapping surveys that cover different parts of the sky (see Paper\,I for further details). 
For the purpose of this study, we mostly focus on studying the orientation of the spiral arms over large scales, as seen in the Cepheid distribution.

\begin{figure}
    \centering
    \includegraphics[width=0.49\textwidth]{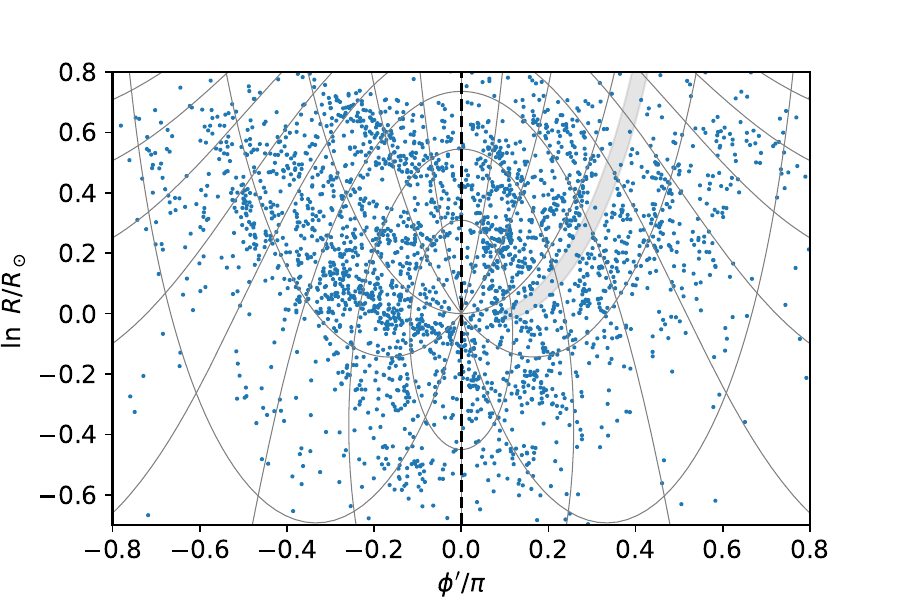}
    \caption{Distribution of dynamically young Cepheids in $\ln R/R_\odot$ (natural log) and $\phi^\prime$, where $\phi^\prime = 0$ is in the direction of the Galactic anti-centre and given in radians. The grey curves show the same polar grid as is overplotted in Fig.~\ref{fig:XY}, and the Cygnus X shadow cone is again shown in the shaded region.
    }
    \label{fig:lnRphi}
\end{figure}

We reconsidered the spatial distribution of the Cepheids in the $\ln R/R_\odot$ and $\phi^\prime$ plane (see Fig.~\ref{fig:lnRphi}), where $\phi^\prime = \pi - \phi$ in radians, so that $\phi^\prime = 0$ is in the direction of the Galactic anti-centre and is positive in the direction of Galactic rotation.  In these coordinates, logarithmic spiral arms should appear as linear features with negative slopes. For $\phi^\prime < 0$ (the third and fourth Galactic quadrants), two such linear features are readily apparent, but not for $\phi^\prime > 0$ (the first and second Galactic quadrants). To confirm and measure in a quantifiable way the orientation (pitch angle) of any spiral arms that might be traced by the Cepheids, we considered a range of possible pitch angles. We defined the rotated coordinate in the $\ln (R/R_\odot)$ -- $\phi^\prime$ plane as 
\begin{equation}
    y^\prime = \ln(R/R_\odot) \cos{\theta} + \phi^\prime \sin{\theta} ,
    \label{eq:yprime}
\end{equation}
where the angle $\theta$ corresponds to an assumed pitch angle, and $y^\prime = 0$ is the $y^\prime$ position of the Sun.  If a spiral arm is present, then we expect that the distribution of Cepheids in $y^\prime$ clearly peaks at the position of the arm when $\theta$ corresponds to the actual pitch angle of the arm. Away from the correct pitch angle for an arm, the peak in the $y^\prime$ distribution becomes broader and less prominent. 

\begin{figure}
    \centering
    \includegraphics[width=0.49\textwidth]{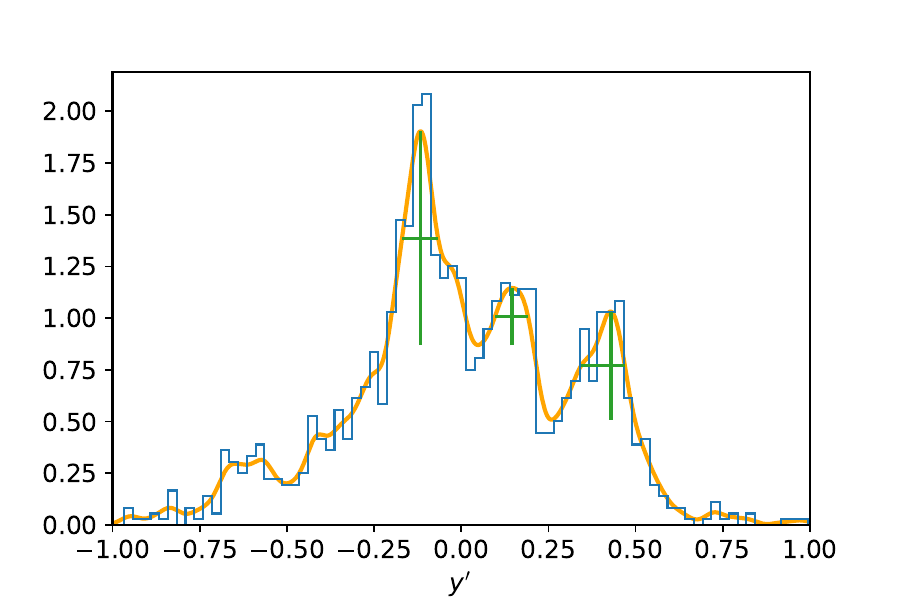}
    \caption{Distribution of dynamically young Cepheids in $y^\prime$ for an assumed pitch angle of 15\deg\ for the sample between azimuths $-120\deg<\phi^\prime<0\deg$. The orange curve shows the KDE density for a bandwidth of 0.025. The three identified peaks are indicated by their measured prominence (vertical green lines) and widths (horizontal green lines). 
    }
    \label{fig:yprime_dist}
\end{figure}

As an example, we show in Fig.~\ref{fig:yprime_dist} the $y^\prime$ distribution of 1431 dynamically young Cepheids in the $\phi^\prime$ range $[-120\deg,0\deg]$ (corresponding to $180\deg <\phi<300\deg$) for an assumed pitch angle of $\theta = 15\deg$. Three clear peaks are seen in the distribution, one inside the position of the Sun ($y^\prime < 0$), and two outside of it. We identified and measured the position of the peaks by first performing a 
kernal density estimate (KDE) using a Gaussian kernel (with a bandwidth of 0.025), as implemented in the {\tt KernelDensity} function from the Python \textsc{sklearn} package \cite{scikit-learn}.  This density was then used as input to the Python {\tt scipy.signal.find\_peaks} function, requiring a minimum peak width of 0.05 (in $y^\prime$) and a minimum peak prominence (i.e. height) of 0.2. The {\tt find\_peaks} function identifies the position of the peaks and gives a measure of the peak prominence and peak width, which is evaluated at half of the relative peak height. 

\begin{figure}
    \centering
    \includegraphics[width=0.49\textwidth]{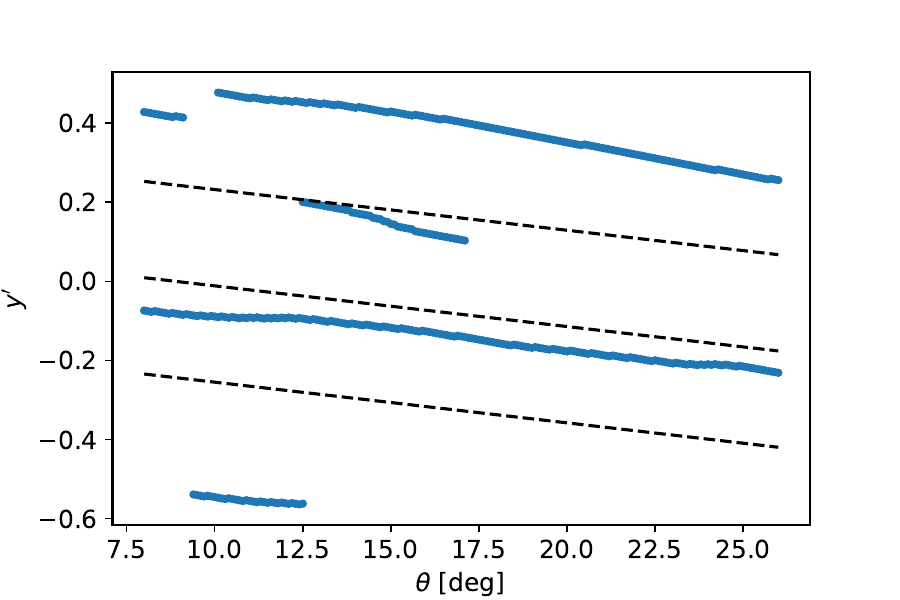}
    \caption{Position of the detected peaks for assumed pitch angles of $8<\theta<26\deg$ for the Cepheids with azimuths $-120\deg<\phi^\prime<0\deg$. The dashed lines show the boundaries we used to assign peak properties to specific putative arms. 
    }
    \label{fig:ypeaks}
\end{figure}

Depending on the azimuth range and assumed pitch angle, the number of peaks ranges from two to four, and one or two arms are generally found within the position of the Sun. The peak positions were used to assign the peak properties (prominence and strength) to a putative arm.  The first arm (at $y^\prime \approx -0.15$ in Fig.~\ref{fig:ypeaks}) inside the position of the Sun, which we identified above as the Sgr-Car arm, is always detected, except for assumed pitch angles $\theta > 22\deg$ for the $\phi^\prime$ range [-30\deg,60\deg]. Another ubiquitous arm is the Perseus arm, the outermost arm detected outside the position of the Sun. We identified this arm because it passes through a large group of Cepheids about 2-3.5\kpc away between galactic longitudes $110\deg<l<150\deg$ when extrapolated into the second quadrant, which is an area with active star formation that has long been identified as the Perseus arm \citep[][]{Morgan1953,vandHulst1954}.  It was also mapped and identified more recently as the Perseus arm with masers by \citet[][hereafter R19]{Reid2019} and in ionised gas by \citet{Haffner1999}.  

Two other arms are sporadically detected. One arm is located inside the Sgr-Car arm, which we tentatively identify as the Scutum arm based on its distance towards the Galactic Centre, and the other arm is located just outside the position of the Sun, but closer than the Perseus arm, which we identify as the Local or Orion arm. We discuss the geometry and identification of the arms in more detail in Sect. \ref{sec:discuss}. 

As mentioned above, we expect the peak corresponding to a spiral arm to be most prominent and narrowest when the angle $\theta$ corresponds to the actual pitch angle of the arm. We therefore define the quantity strength as the 
peak prominence divided by peak width. As the distribution in azimuth of our sample is not centred at $\phi^\prime = 0$, we find that the peak positions in general drift to lower $y^\prime$ values as the angle $\theta$ is increased. Figure~\ref{fig:ypeaks} shows for the dynamically young Cepheids in the $\phi^\prime$ range $[-120\deg,0\deg]$ the position of the detected peaks as we consider possible pitch angles between 8\deg\ and 26\deg\ in steps of 0.1\deg. The dashed lines show the boundary criteria we used to assign the peak properties to specific arms. Any peaks above/below the upper/lower boundaries are assigned to outer/inner-most arms, allowing for the detection of up to four arms.  The boundary between the outer and inner arms (middle dashed line) is 
\begin{equation}
    y^\prime_b = 0.4 \left(\, \langle \ln(R/R_\odot) \rangle \cos{\theta} + \langle \phi^\prime \rangle \sin{\theta}\, \right) - 0.3\, \theta,
    \label{arm_boundary}
\end{equation}
where the quantities between the angular brackets are the median values of the selected sample, and $\theta$ is in radians.  The first term sets the vertical offset in $y^\prime$, and the second term primarily determines the slope of the lines. 
The coefficients 0.4 and 0.3 were found by trial and error, verifying that they were sufficient for all the different azimuth ranges we considered. The other two boundaries were set at $\pm \Delta y^\prime$ from $y^\prime_b$, taking $\Delta y^\prime$ as half of the distance between the outermost peak and the first peak inside the position of the Sun, that is, the peaks corresponding to the Perseus and Sgr-Car arms. In the rare case when no outer peak was detected, the default interval in $\Delta y^\prime = 0.25$ was used.  

\begin{figure}
    \centering
    \includegraphics[width=0.49\textwidth]{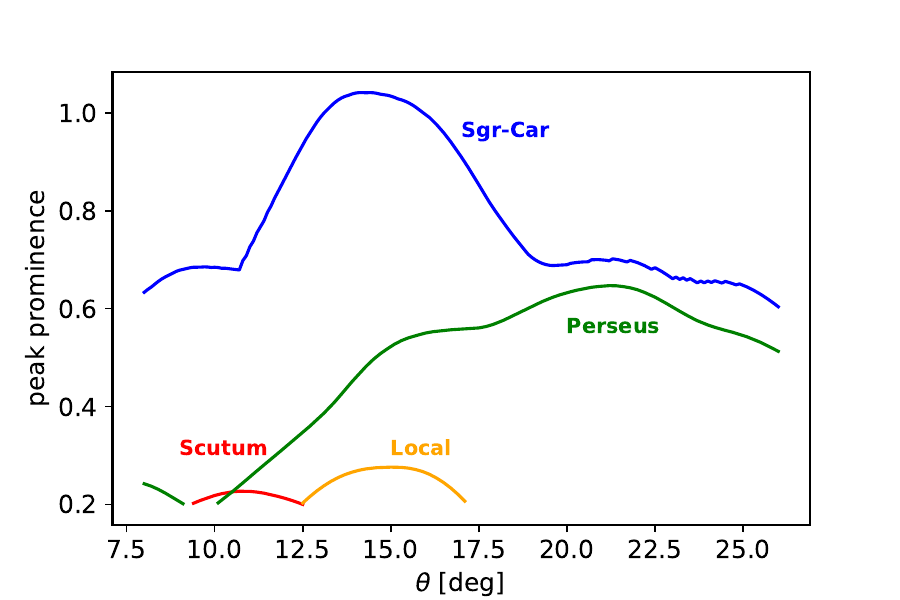}
    \includegraphics[width=0.49\textwidth]{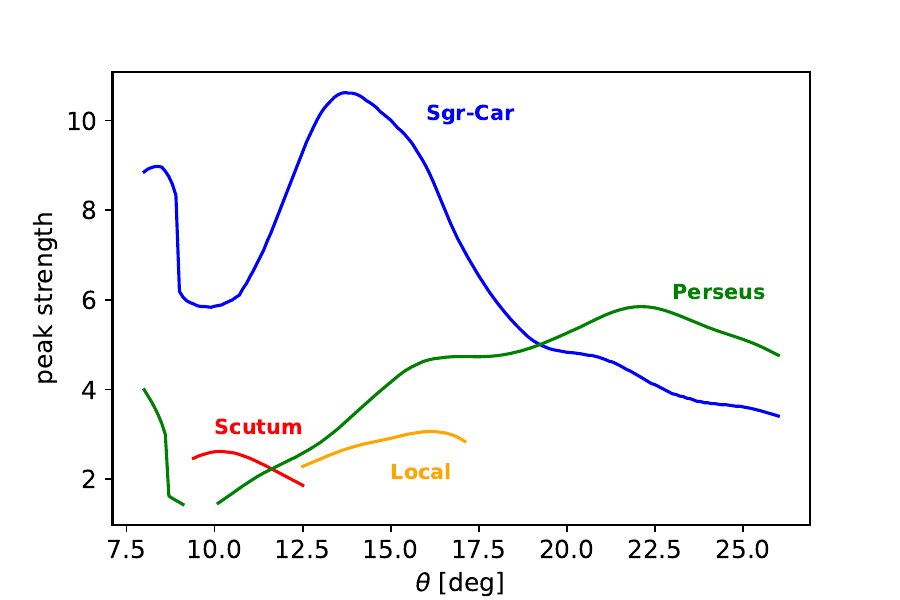}
    \caption{Variation in the prominence and strength of the peaks with respect to the assumed pitch angle $\theta$ for the Cepheids with azimuths $-120\deg<\phi^\prime<0\deg$. The blue curve shows the first arm inside the position of the Sun (Sgr-Car), the gold curve shows the first arm outside this position (the local Orion arm), and the green curve shows the farthest arm we detected.  
    }
    \label{fig:prom_strength}
\end{figure}

After the peak properties were assigned to each arm for different possible pitch angles $8<\theta<26\deg$, we determined the angles at which the peak prominence and peak strength is maximum, taking these angles as the two possible pitch angles of the arm. While for most azimuth ranges and data selections there is only one clear maximum over the range of possible pitch angles, Fig.~\ref{fig:prom_strength} shows that the peak strength might show two maxima.  This behaviour in the peak strength arises in part because the strength is more sensitive to weaker but narrower peaks in the $y^\prime$ distribution. The maximum in the peak strength for the Sgr-Car occurs at nearly the same pitch angle $\theta$ as the maximum in the peak prominence.   

In summary, for the data selection used in this section (dynamically young Cepheids over the $\phi^\prime$ azimuth range of [-120\deg,0\deg]), we find pitch angles of 10.7, 14.2, 15, and 21.3\deg for the four detected arms when we considered the peak prominence, and $10.1, 13.7, 16.1, 22.1\deg$ for the peak strength. Fig.~\ref{fig:prom_strength} shows that the Scutum and Local (Orion) arm are quite weak with respect to the Sgr-Car and Perseus arms. Nevertheless, the estimated pitch angles for each arm fall within a couple of degrees of each other for all the arms. We also note that the pitch angles of the arms increase with distance outward from the Galactic centre. 

\begin{figure}
    \centering
    \includegraphics[width=0.49\textwidth]{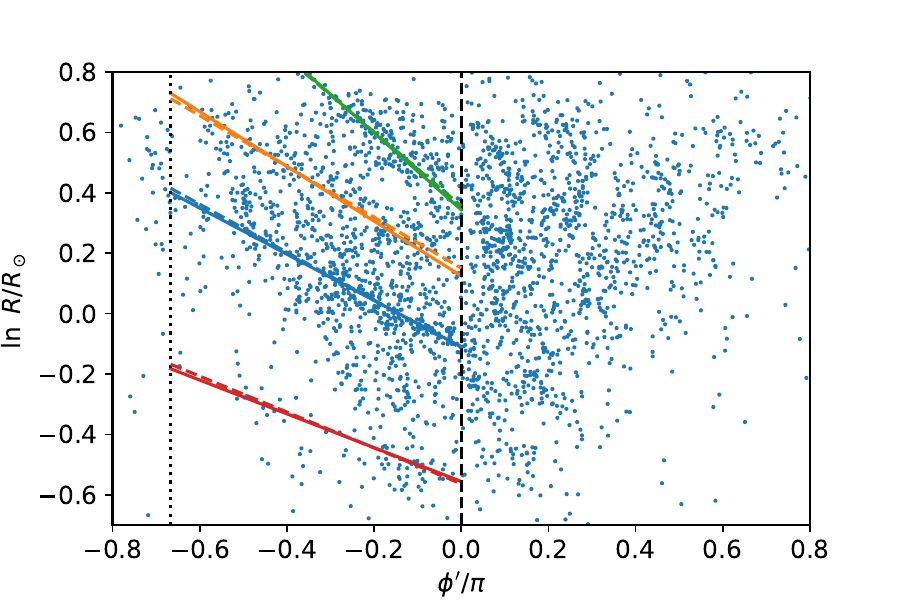}
    \caption{Distribution of young DR3 Cepheids in $\log R/R_\odot$ and $\phi^\prime$, as in Fig.~\ref{fig:lnRphi}, but the positions of the arms are overplotted as derived using the Cepheids with azimuths $-120\deg < \phi^\prime < 0\deg$. Red corresponds to the Scutum arm, blue to the Sgr-Car arm, orange the local Orion arm, and green the outermost (Perseus) arm. The solid (dashed) lines show the arms using the pitch angles found from the maximum strength (prominence).}
    \label{fig:lnRphi_arms}
\end{figure}

After a pitch angle for an arm was found, we determined the position of the arm using the $y^\prime$ position of the peak at that pitch angle in Eq. \ref{eq:yprime}, taking $\phi^\prime=0$ and solving for $\ln(R)_{\phi^\prime=0}$, which we designate as $\ln(R_0)$.  A spiral arm
described by the equation 
\begin{equation}
\label{eq:spiRal}
    R = R_0 e^{-\tan\theta\, \phi^\prime}
\end{equation}
then corresponds to the equation
\begin{equation}
    \ln (R/R_\odot) = ln(R_0/R_\odot) - \phi^\prime \tan\theta\, .
\end{equation}
Using this equation, we overplotted the derived position of the arms on the $\ln R$--$\phi^\prime$ distribution, as shown in Fig.~\ref{fig:lnRphi_arms}.) 
Each spiral arm is thus characterised by two parameters: the pitch angle $\theta$, and $\ln(R_0)$, which are derived from the pitch angle and the measured $y^\prime$ position of the peaks. 

We estimated the uncertainty of the measured peak positions by performing a bootstrap of $N=100$ samples of the $y^\prime$ distribution. Using the standard deviation of the peak positions, we find that the uncertainty of the peak positions for the three detected arms (at $\theta = 15\deg$ and for the sample Cepheids between $-120\deg<\phi^\prime<0\deg$, shown in Fig.~\ref{fig:yprime_dist}) to be 0.005, 0.019, and 0.017, showing that the peak positions are well determined. 
The same bootstrapping technique was also used to test the robustness of the detection of each arm. For instance, at a pitch angle of $\theta = 15\deg$, the Sgr-Car and Perseus arm was found in all 100 bootstraps, while the Local (Orion) arm was only detected 86\% of the time.  We repeated this procedure for a pitch angle of $\theta = 11\deg$ to allow us to detect the Scutum arm and found uncertainties in the $y^\prime$ positions of the peaks of 0.026, 0.009, 0.048, and 0.028.  
Again, for 100 bootstraps the Sgr-Car arm is always detected, while the detection rates for the Scutum arm is 74\%, for the Local arm is only 41\%, and that of the Perseus arm is 88\%. That the Perseus arm is not always detected at this small pitch angle is expected, while the sporadic detection of the Scutum and Orion arms even at an angle $\theta$ near their pitch angles indicates how weak they are.

The resulting uncertainties of the determined pitch angles are more difficult to estimate because they depend on the variation in the $y^\prime$ distribution when $\theta$ is varied, which is determined by a given set of measured Cepheid positions in the Galactic plane. The uncertainty of these positions in turn are a consequence of our distance uncertainties. We generated alternative sets of measured Cepheid positions by assuming that the distance uncertainties are well described by a Gaussian distribution in the distance modulus of each Cepheid with a standard deviation of $\sigma_\mu$, the uncertainty in the distance modulus. In this way, we generated 100 samples from our dataset. We analysed each sample as described above, identifying the peaks in the $y^\prime$ coordinate for a range of possible pitch angles, measuring their prominence and strength, and identifying the pitch angle for each arm for which these properties are maximum.   We took the standard deviation of our 100 determinations of the pitch angle for each arm as an estimate of the uncertainty of the pitch angle. Using the peak prominence, we find uncertainties of 1.0, 0.7, 1.7, and 1.4\deg\ for the Scutum, Sgr-Car, Orion, and Perseus arms, respectively. Similarly, using the peak strength, we find uncertainties of 1.2, 1.4, 1.8, and 3.2\deg. As expected, the estimated pitch angles from the two methods agree with each other within the uncertainties. 

From the same 100 samples, we also measured the robustness of the detection of each arm, similarly as was done above when we bootstrapped the $y^\prime$ distribution for our dataset. We find that the Sgr-Car and Perseus arms are detected in 100\% of the resamples, the Local (Orion) arm is detected in 74\%, and the Scutum arm is detected in only 32\% of the resamples.  

  
\section{Results}

\begin{table*}[ht]
\caption{Spiral arm parameters for each arm over different azimuth ranges. }
\label{tab:arm_attributes}
\centering
{ \footnotesize
\begin{tabular}{lccccccccccccccccc}
\hline\hline
  &  &  \multicolumn{4}{c}{Scutum} &  \multicolumn{4}{c}{Sgr-Car} & \multicolumn{4}{c}{Orion} & \multicolumn{4}{c}{Perseus} \\
  &  &  \multicolumn{2}{c}{strength} & \multicolumn{2}{c}{prominence}  &  \multicolumn{2}{c}{strength} & \multicolumn{2}{c}{prominence}  & \multicolumn{2}{c}{strength} & \multicolumn{2}{c}{prominence} & \multicolumn{2}{c}{strength} & \multicolumn{2}{c}{prominence} \\
$\phi$ range & N & $\theta$ & $\ln R_0$ & $\theta$ & $\ln R_0$ & $\theta$ & $\ln R_0$ & $\theta$ & $\ln R_0$ & $\theta$ & $\ln R_0$ & $\theta$ & $\ln R_0$ & $\theta$ & $\ln R_0$ & $\theta$ & $\ln R_0$ \\
\hline
$[-120,0]$ & 1431 & 10.1 & 1.56 & 10.7 & 1.55 & 13.7 & 2.0 & 14.2 & 2.0 & 16.1 & 2.24 & 15.0 & 2.26 & 22.1 & 2.46 & 21.3 & 2.47 \\
$[-90,0]$ & 1331 & 10.2 & 1.56 & 10.6 & 1.55 & 14.5 & 2.0 & 14.8 & 2.0 & 16.0 & 2.24 & 15.6 & 2.25 & 22.0 & 2.46 & 20.9 & 2.48 \\
$[-120,30]$ & 2012 & --- & --- & --- & --- & 13.4 & 2.01 & 15.7 & 1.99 & 17.3 & 2.21 & 16.7 & 2.22 & 21.6 & 2.47 & 21.2 & 2.47 \\
$[-90,30]$ & 1912 & --- & --- & --- & --- & 14.3 & 2.0 & 15.5 & 1.99 & 17.5 & 2.21 & 17.0 & 2.22 & 21.5 & 2.47 & 20.9 & 2.47 \\
$[-60,30]$ & 1637 & --- & --- & --- & --- & 11.4 & 2.01 & 15.0 & 1.99 & 19.3 & 2.22 & 18.7 & 2.22 & 21.3 & 2.47 & 20.7 & 2.47 \\
$[-120,60]$ & 2422 & --- & --- & --- & --- & 13.6 & 2.01 & 15.7 & 1.99 & 17.9 & 2.2 & 16.9 & 2.21 & 20.2 & 2.48 & 20.8 & 2.48 \\
$[-90,60]$ & 2322 & --- & --- & --- & --- & 14.2 & 2.0 & 15.5 & 1.99 & 18.6 & 2.21 & 17.3 & 2.21 & 20.2 & 2.48 & 20.6 & 2.48 \\
$[-60,60]$ & 2047 & --- & --- & --- & --- & 12.1 & 2.01 & 15.0 & 2.0 & 19.7 & 2.22 & 19.1 & 2.22 & 20.0 & 2.48 & 20.3 & 2.48 \\
$[-30,60]$ & 1518 & --- & --- & --- & --- & 12.0 & 2.01 & 14.7 & 2.0 & 22.7 & 2.25 & 21.5 & 2.24 & 18.4 & 2.47 & 22.5 & 2.47 \\
\hline
\end{tabular}
\tablefoot{$N$ is the number of Cepheids found in the Galactic azimuth $\phi$ range in degrees, the pitch angle $\theta$ is in degrees, and $\ln R_0$ is for $R_0$ in \kpc. }
} 
\end{table*}

In the previous section, we described our method using the 1431 dynamically young Cepheids with azimuths $-120\deg<\phi^\prime<0\deg$ as an example. In this section, we consider alternative selected azimuth ranges to explore the possible extent of each arm as well as possible variation in the pitch angles. We also discuss the effect of alternative age selections.  

We always explored azimuth ranges that are at least 90\deg\ in extent and that started at azimuths $\phi^\prime < 60\deg$ ($\phi > 120\deg)$. Azimuths at $\phi^\prime > 60\deg$ were not considered because we lack data in the inner disk as a consequence of extinction, which limits how far we can trace the arms into the first quadrant. For all the azimuth ranges we considered, we used the same scheme as in the previous section to identify the arms: the first arm within the position of the Sun is identified as Sgr-Car, and the outermost arm as Perseus. 
Table \ref{tab:arm_attributes} summarises our results for each arm for the different azimuth ranges we considered, listing the different azimuth ranges, the number of Cepheids found in each azimuth range, and the spiral arm parameters found using either the peak prominence or peak strength. 

\begin{figure*}
    \centering
    \includegraphics[width=0.49\textwidth]{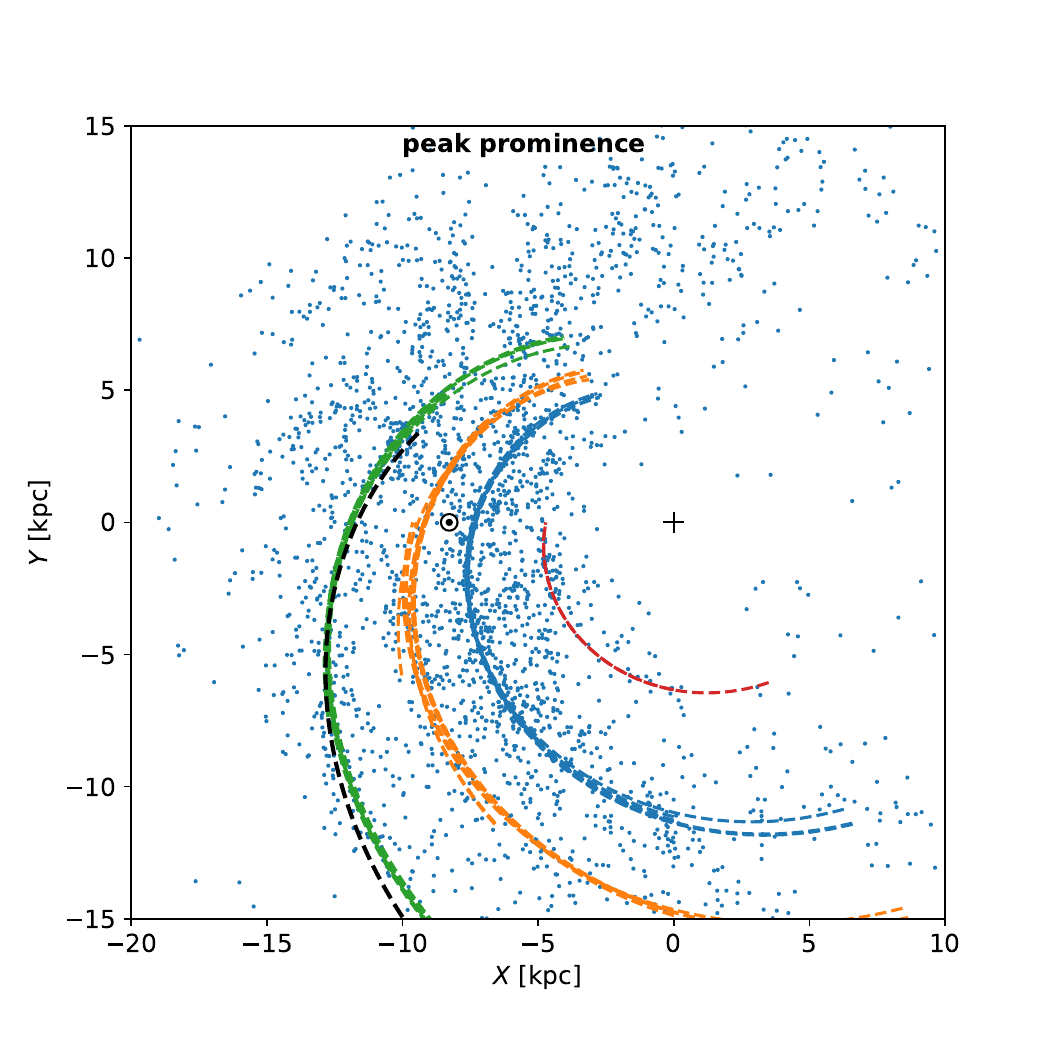}
    \includegraphics[width=0.49\textwidth]{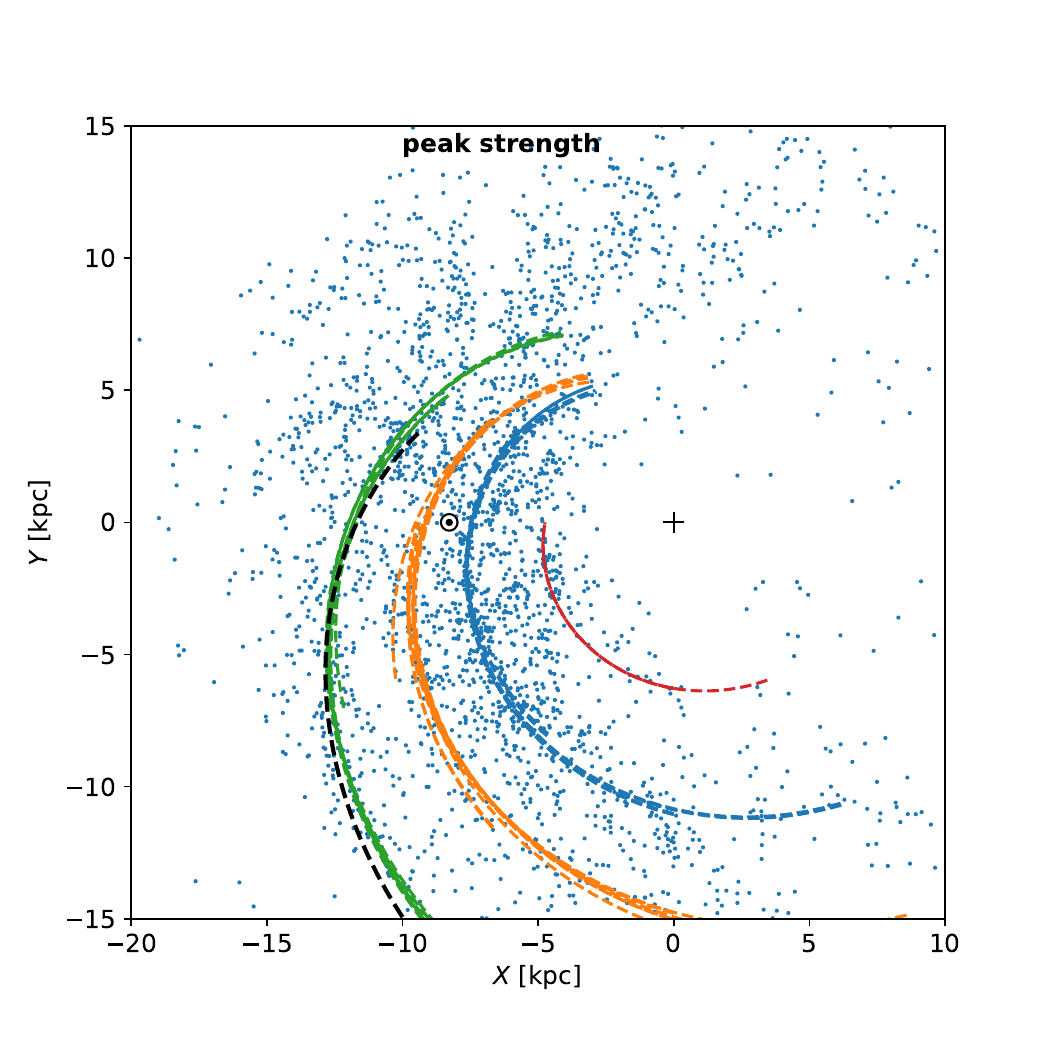}
    \caption{XY distribution of the dynamically young Cepheids, overplotted with the arms from all considered azimuth ranges. The red curve corresponds to the Scutum arm, the blue curve shows the Sagittarius-Carina arm, the gold curve shows the Local (Orion) arm, and the green curve shows the Perseus arm.  The dashed black curve is an arm detected by \cite{Levine06} in the HI. For an interactive version of this plot see \url{https://gaia-unlimited.org/map-of-milky-way-cepheid-variables/}
    }
    \label{fig:XY_arms}
\end{figure*}

We overplot all of the spiral arm fits in Fig.~\ref{fig:XY_arms} for all the azimuth ranges 
on the galactocentric $XY$ positions of the Cepheids using Eq. \ref{eq:spiRal}.
In general, the positions of the arms at $\phi^\prime = 0$ ($\ln R_0$) are very consistent for all the arms in all the azimuth ranges we considered, regardless of whether we used the strength or prominence of the arm to estimate the arm parameters. The coincidence of the arm positions for the different azimuth ranges confirms that using the position of the peaks in $y^\prime$-space with respect to the Sun gives a reliable identification of the arms.
The pitch angles of the arms are also fairly consistent for all the azimuth ranges, with the exception of the Local (Orion) arm, which has larger pitch angles when data from the first and second quadrants ($\phi^\prime > 0.$) are included, increasing from 15\deg\ (similar to the Sgr-Car arm) to 22\deg\ (similar to the Perseus arm). 
The pitch angles of the Perseus arm are quite consistent between the two methods, with a very gradual progression from 22\deg\ to 20\deg\ for azimuth ranges from negative to positive $\phi^\prime$, except for the azimuth range [-30\deg, 60\deg], where the two methods diverge. 
The pitch angle of the Sgr-Car is more consistent over the azimuth ranges considered, with the pitch angles from the arm strength being slightly smaller than those from the arm prominence.
The overplotted spiral arms for all the arms nearly overlap with each other, with the Local (Orion) arm showing the most variation in pitch angle when different azimuth ranges are considered.  

The pitch angles of the arms are not the same, but increase with distance outward from the inner to the outer Galaxy from about 10\deg\ for the Scutum arm to about 20-22\deg\ for the Perseus arm. This trend, noted in the previous section, is preserved for all azimuth ranges considered.

Although the Cepheids in the first and second quadrants do not show obvious spiral arm structure, the clumpy distribution of the Cepheids is apparently consistent with the assumption that the spiral structure in the third and fourth quadrants extends into the first and second quadrants, at least for the three outermost arms. In contrast, the innermost Scutum arm is not detected when data from the first and second quadrants is included. Indeed, the detection of this arm is quite weak, and we can only claim a tentative detection of this arm.  
The Local (Orion) arm, on the other hand, is detected at all azimuth ranges, but it is always weaker than the Sgr-Car and Perseus arms.  

As the individual Cepheid ages are very uncertain, we explored the effect of alternative age selections. We first repeated the analysis for all the azimuth ranges with the selection
\begin{equation}
    Age_{Myr} < \frac{\pi R}{0.2314}\,, 
    \label{new_age_crit}
\end{equation}
that is, requiring the Cepheids to be younger than half the dynamical time (Galactic rotation period) at their current galactocentric radius.  This is a more stringent selection than the one used above and reduced our sample to 1894 Cepheids. With this selection, there are still sufficient Cepheids in the inner disk, but most of Cepheids in the outer disk are lost. This means that most of the advantage of using a dynamical age selection is lost as well.  As a consequence, our detection of the outer (Perseus) arm is much weaker.  Nevertheless, it is still detected with pitch angles similar to those reported above, even though the peaks in strength and prominence are far weaker.  The Local (Orion) arm is also still detected, but the pitch angles vary more strongly, with larger systematic differences between the two methods. The Sgr-Car arm remains easily detected, while the Scutum arm is just barely detected, but both arms have similar pitch angles as reported above. 

When we instead impose a simple age cut of $Age_{Myr} < 200$~Myr, then the Perseus arm is only very weakly detected for the two azimuth ranges restricted to the third and fourth quadrant, without a clear peak with respect to pitch angle in either the arm strength or prominence, while the Local (Orion) arm is no longer detected. For the other azimuth ranges that include Cepheids in the first and second quadrant, both arms show the same trends in pitch angles as reported above. The Sgr-Car and Scutum arms are unaffected, however, and show the same pitch angles as reported above.

\section{Discussion}
\label{sec:discuss}

In the previous section, we derived spiral arm parameters over a suite of overlapping azimuth ranges (See Table \ref{tab:arm_attributes}.) Although the arm parameters for a given arm do not vary by a large amount, this raises the question of which set of parameters should be used for each arm. The spiral structure in the Cepheid distribution is most clearly evident in the third and fourth quadrants. We therefore suggest to use the set of arm parameters in the azimuth range $[-90\deg, 0\deg]$, where the pitch angles derived from both methods (strength or prominence) agree quite well for all the arms. Averaging the pitch angles and $\ln R_0$ from the two methods in this azimuth range results in the recommended spiral arm parameters given in Table \ref{tab:arm_params}.  The uncertainties were estimated by resampling the distances 100 times, redetermining the arm attributes for each resample, and taking the standard deviation of the arm attributes as the uncertainty, as described at the end of Sect. \ref{sec:spirals}. The uncertainties quoted in the table are those based on the arm strength, which for the pitch angles are larger by about a factor of two than those found using the arm prominence for the Carina and Perseus arms. 

As already noted in Sect. \ref{sec:spirals}, the pitch angles of the arms increase from the inner to the outer arms. This progression would apparently argue against these arms being part of a common grand-design spiral pattern because they do not share the same pitch angle. However, we cannot help but note that the logarithmic radial separation between the Scutum and Sgr-Car arms ($\Delta  \ln R_0 = 0.44$) is almost the same as that between the Sgr-Car and Perseus arms (0.46).  (Taking 0.45 as the $\Delta  \ln R_0$ separation between a four-arm spiral pattern would imply a pitch angle of about 16\deg.) While suggestive, conclusions drawn from these arm spacings can only be tentative because the detection of the Scutum arm is itself tentative, and even if it is a positive detection, its distance may be biased given the uncertainties in the extinction at this direction and distance. 

\begin{figure*}
    \centering
     \includegraphics[width=0.49\textwidth]{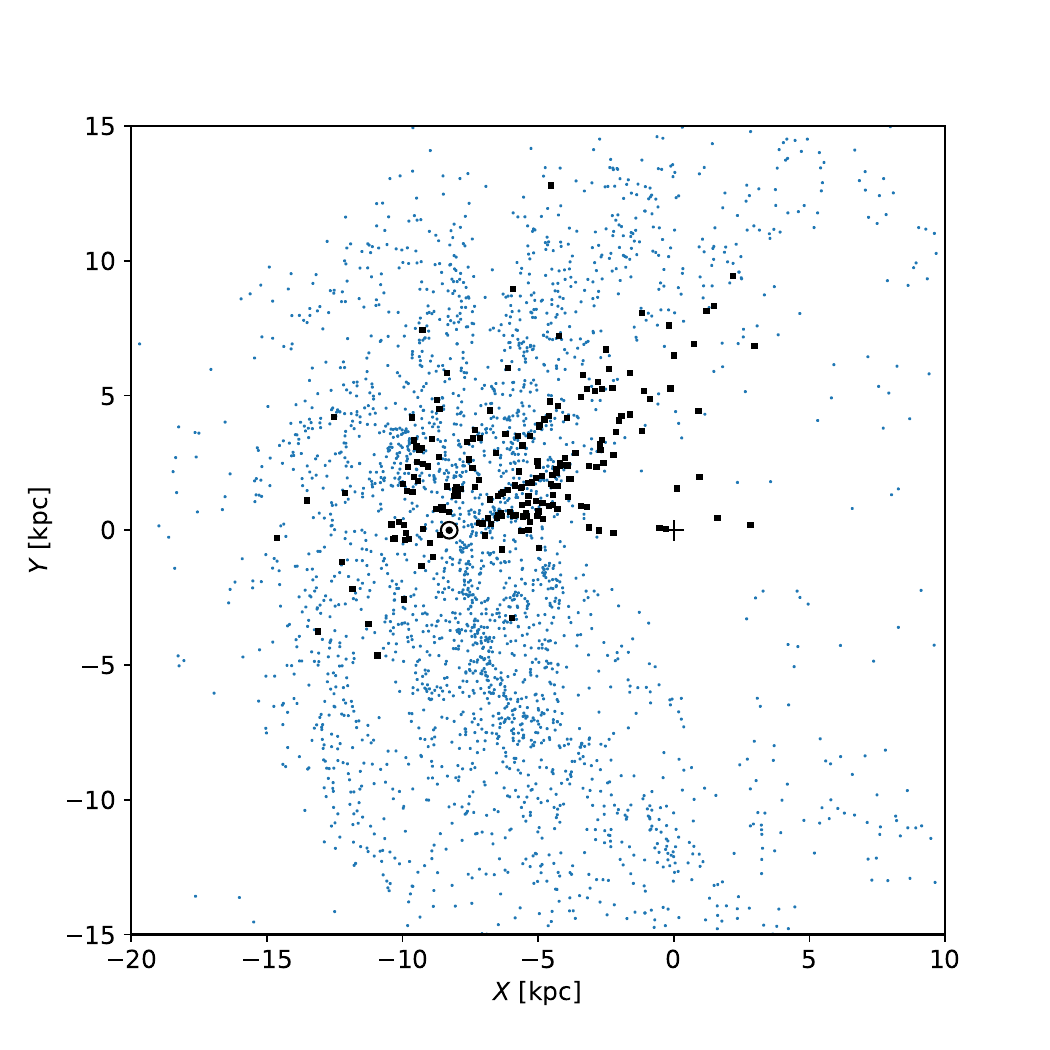}
    \includegraphics[width=0.49\textwidth]{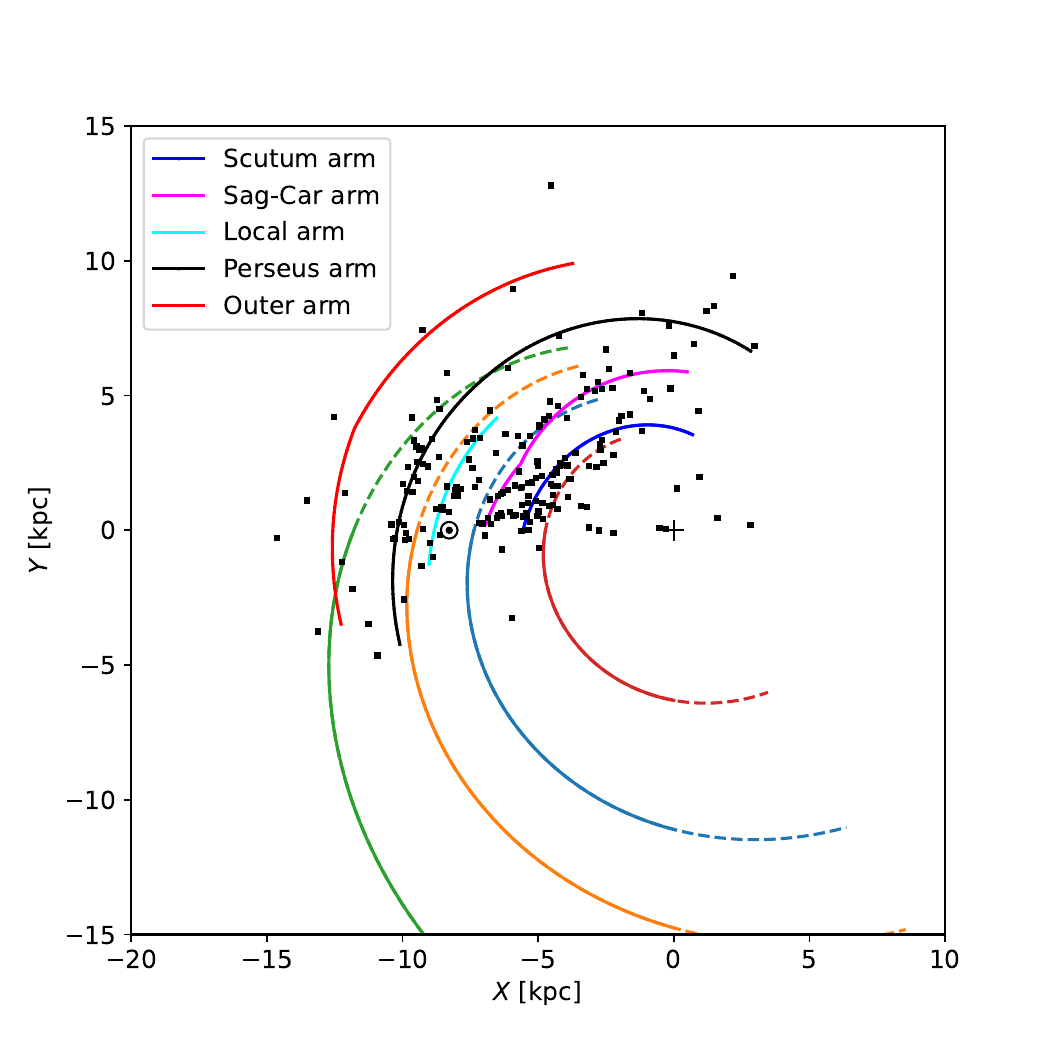}
    \caption{Comparison between Cepheids and maser sources. Left panel: 
    Comparison of the Cepheid sample (blue dots) with the masers from R19, shown as black squares. Right panel: Comparison between the spiral model derived in this work (the colours of the spiral arms are the same as in previous figures) and the spiral arm model from R19 based on the masers.
    }
    \label{fig:comparison_reid}
\end{figure*}

\begin{table}[ht]
\caption{Recommended spiral arm parameters }
\label{tab:arm_params}
\centering
\begin{tabular}{lcc}
\hline\hline
Arm & $\theta$ & $\ln R_0$ \\
\hline
Scutum &  $10.4 \pm 1.4 $ & $1.56 \pm 0.026$ \\
Sgr-Car &  $14.7 \pm 1.2$ & $2.0 \pm 0.008$ \\
Orion &  $15.8 \pm 2.1$ & $2.25 \pm 0.028$ \\
Perseus &  $21.5 \pm 3.1$ & $2.47 \pm 0.04$ \\
\hline
\end{tabular}
\tablefoot{Based on the Cepheid distribution in the third and fourth quadrants in the $\phi^\prime$ azimuth range $[-90\deg, 0\deg]$.  The pitch angle $\theta$ is in degrees, and $\ln R_0$ is for $R_0$ in \kpc.  }
\end{table}

The most obvious spiral arm in the Cepheid distribution is the Sgr-Car arm in the fourth quadrant. For this arm, we note that not only do we find consistent pitch angles for all azimuth ranges considered, but also a tangent of this arm in the first quadrant at $l \approx 50\deg$, which has historically been associated with the Sagittarius arm.  
(For a detailed discussion of the history of the Sagittarius arm, see Appendix A of \citet{Kuhn:2021}.)
That is, regardless of whether we extrapolate the spiral arm derived from the Cepheids in the fourth Galactic quadrant or also use the Cepheids in the first quadrant, the nearest arm inside the solar circle has arm tangents that correspond to both the Carina arm (at $l \approx 285\deg$) and the Sagittarius arm (at $l \approx 50\deg$), suggesting that these two tangents are indeed from a single arm that has been traditionally recognised as the Sagittarius-Carina (Sgr-Car) arm. 

The second most evident arm in the Cepheid distribution is the outermost Perseus arm in the third quadrant, which we find to coincide with one of the four arms reported by \citet{Levine06} (their arm 2; see their Table 1 and Fig.~4A) in HI, shown in Fig.~\ref{fig:XY_arms} as the dashed black curve, consistent with previous findings \citep{Poggio2021,Drimmel23_gaia}. According to their location, two other arms of Levine (their arms 3 and 4) coincide with the Local and Sagittarius arm in the fourth quadrant, but these have significantly  different orientations than those seen here in the Cepheids. 

In contrast, the Local (Orion) arm is much weaker than either the Sgr-Car or the Perseus arm. 
Indeed, we may only be detecting it because our position is favourable with respect to this arm. 
Nevertheless, it is evident over a wide range of azimuths, confirming that it is a real feature of significant length, as was recently suggested \citep{Xu2016, Xu:2018, Poggio2021}. However, a slightly different orientation of the Local arm was suggested before: 
Using \gaia\ DR2 data, \citet{Xu:2018} suggest that the Local arm continues into the fourth quadrant and bends inward toward the Galactic centre. On the other hand, the Local arm from the overdensity maps of upper main-sequence stars from \gaia{} DR3 in \citet{Poggio2021} appears to extend from the first to the third quadrant, with a more open geometry. This agrees with the orientation of the Local arm seen here in the Galactic Cepheids. 
Judging from the distribution of the Cepheids, it is certainly not one of the major arms of the Milky Way, but while it is much weaker, it seems to be as long as the Sgr-Car arm.   

It is worth comparing our results with other recent studies of Galactic spiral structure based on Cepheids. 
\citet{Bobylev2022} used about 600 pre-selected Cepheids from \citet{Skowron2019a,Skowron2019b} to measure the pitch angle of the Sgr-Car arm and an outer arm beyond Perseus.  Performing a linear least-squares fit in $\ln(R) - \phi$ space, they reported similar pitch angles for both arms.  For the Sgr-Car arm, which is most clearly seen in their dataset, they found a pitch angle near 12\deg.  However, their solution for each arm is already constrained by their pre-selection over a limited range in $\ln(R) - \phi$. 
In contrast, our method does not rely on pre-assigning Cepheids to specific arms that are identified by the peaks in the distribution of the Cepheids in $\ln(R) - \phi$ space.  

\citet{Lemasle2022} derived distances using MIR Wesenheit magnitudes, which may still show extinction effects, as discussed in Paper I. To identify the spiral structure, they 
used the t-SNE+HDBSCAN algorithm, feeding the HDBSCAN clustering algorithm \citep{Campello2013} the output from the two-dimensional 
t-distributed Stochastic Neighbor Embedding (t-SNE, \citet{vandermaaten2008}), 
generated from providing the algorithm the $(\phi,\ln R)$ coordinates of the Cepheids. Primarily based on the subsample of Cepheids younger than 150~Myr, they identified numerous spiral arm segments. They noted, however,
that the algorithm is sensitive to small gaps causing a given spiral arm to be split in several segments.  They then reported about 18 spiral segments. This result should not be interpreted as evidence that the Milky Way is flocculent: Spiral arms on kiloparsec scales are not continuous structures, but are a perceived pattern on large scales that consists of spurs, feathers, and large disconnected star formation regions. In addition, extinction effects can introduce additional apparent gaps that are not real. For these reasons, we applied the logarithmic spiral formalism only to datasets covering a wide range ($\Delta \phi \geq 90\deg$) in Galactic azimuth to map the spiral structure on a large scale.

The most constraining data for the large-scale structure of the spiral arms to date were the masers found in high-mass star-forming regions \citep[][hereafter R19]{Reid2019}.
The Sgr-Car arm was identified with the well-kmown tangent at $l \approx 283\deg$ \citep{Bronfman:2000}. Using this tangent together with masers in the first quadrant, R19 fit a log-periodic spiral, allowing for a kink in the arm, with different pitch angles on either side of the kink. In this way, R19 connected both arm tangents to construct the Sgr-Car arm and obtained a pitch angle of $17.1 \pm 1.6\deg$ after the kink (i.e. towards the fourth quadrant) and a pitch angle of $1.0 \pm 2.1\deg$ before the kink (i.e. in the direction of Galactic rotation). However, for the Cepheids, we find that the Sgr-Car arm geometry that we deduce from the Cepheids, with a pitch angle of about 15\deg, is able to account for both arm tangents without the need of a kink or additional constraints, and it also passes through the masers that are located in the direction of the arm tangent at $l \approx 50\deg$. 

While the large-scale Cepheid distribution seems to support the geometry of the Sgr-Car arm that is traditionally assumed, we note that the position of the Sgr-Car kink employed by R19 roughly corresponds with a gap observed in the maser distribution, which we also see in the Cepheid distribution. \cite{Kuhn:2021} identified 25 star-forming regions in the galactic longitude range $l \approx  4.0 \degree - 18.5 \degree$ that are arranged in a long linear structure with a high pitch angle of 56$\degree$. We observe a similar structure in our Cepheid sample and note that it is also present in the maser distribution. One possible interpretation is that this high pitch-angle structure is a spur that departs from the Sgr-Car arm. When we inspect external spiral galaxies, spurs (luminous features) and feathers (dust features) that extend from spiral arms to inter-arm regions are frequent. An alternative interpretation is that this high pitch-angle structure represents the main spiral arm itself, so that the arm abruptly changes in pitch angle in the first quadrant. In this scenario, the gap in the stellar distribution observed in the Cepheids and masers would represent an inter-arm region, and the two tangents at $l \approx 285\deg$ and $l \approx 50\deg$ would not be part of the same arm. 

The inner regions contain a small azimuthal range that is covered by both our model and the R19 model for the Scutum arm. Fig.~\ref{fig:comparison_reid} (right panel) shows that the Scutum arm derived here has a very similar pitch angle to the arm from R19, but slightly shifted toward the inner parts of the Galaxy.  The Orion (Local) arm in the Cepheids agrees well with the arm seen in the masers, but we find a somewhat larger pitch angle than was derived by R19 ($11.4\degree$). 

In contrast with the other arms, the orientation of the Perseus arm in the two models is entirely different. As already mentioned, the geometry of this arm in the Cepheids agrees quite well with an arm seen in HI as well as with young upper-mainsequence stars in the third quadrant. Accepting the R19 mapping of the Perseus arm in the second quadrant, we must conclude that either the Perseus arm changes abruptly in pitch angle, or that the Cassiopeia region is the end of the Perseus arm and a different set of arms dominates the outer disk, and the Cassiopeia star formation region is located where these two arms intersect. 

When we consider the source distribution alone and ignore the models, the positions of Cepheids and masers in the Cassiopeia region (the overdensity at l $\approx 110-140 \degree$) do not perfectly coincide: The masers are slightly shifted toward lower galactocentric radii compared to the distribution of Cepheids. Based on the distance determinations obtained in Paper I and this work, we conclude that this shift is not due to distance errors.
Offsets between the distribution of masers and Cepheids, seen here in the Perseus arm and also in the Scutum arm, could be explained by the age difference between the two populations and the difference between the angular pattern speed of the arm and angular rotation rate of the stars: Maser emission originates from circumstellar material around high-mass stars and is therefore expected to be very young, while the Cepheids we selected are older (see Fig.~\ref{fig:ages}) and therefore had time to move away from their birthplaces. The spiral structure they form may therefore be different than the one formed by masers (see Fig.~3 in \citealt{Skowron2019a}). 
However, the offset seen for the Perseus arm would require this section of the arm to be inside its corotation, while the offset seen in the Scutum arm would require the Cepheids in this arm to be outside its corotation.  

While part of the observed differences between the Cepheid distribution and the masers might be real and physically motivated, it is important to note that the comparison between our results and the R19 model strongly depends on the criterion adopted to assign each maser to a spiral arm. For instance, if the nearby masers at $l \approx 180 \degree$ were assigned to the Local arm instead of being assigned to the Perseus arm (as was done in R19), the resulting geometry of the Perseus and Local arms would be different and would presumably agree better with the geometry obtained here. In any event, we also tried to apply our method to the maser sample, but found that it did not give reliable results due to the lack of sources, as the maser sample only numbers about 200 and covers a smaller range of Galactic azimuth.

\section{Conclusions}
\label{sum}
We have mapped the large-scale spiral structure of the Galaxy with dynamically young Cepheids, using new distance estimates based on MIR photometry from an accompanying paper (see Paper I) and a new age criterion for selecting young Cepheids.  Our method for deriving spiral arm parameters does not rely on pre-assigning sources to specific arms, but instead detects and derives arm parameters based on the overall distribution of Cepheids over a wide range in Galactic azimuth. Our approach was informed by the understanding that the logarithmic spiral originates as an empirically motivated geometrical model to describe the morphology of disk galaxies on large scales, and that spiral arms are composed of a rich variety of discontinuous substructure on the kiloparsec scale that are often not well described by a logarithmic spiral. In addition, for our own Galaxy, the additional difficulty arises that we lack an external global view of the disk, but our samples of spiral tracers are unavoidably incomplete and limited by extinction, which introduces additional gaps in our knowledge of the true distribution of sources. For the Cepheids, this is especially an issue in the parts of the first quadrant that do not fall within the survey footprint of the Optical Gravitational Lensing Experiment (OGLE). Nevertheless, Cepheids are an excellent tracer of the young stellar population.

It is important to note that our source list stems from a compilation of various surveys \citep{Pietrukowicz:2021} that differed in their on-sky footprint, their cadence, photometric bands, magnitude limits, and so on (see Fig.~2 of Paper I.) This introduces inhomogeneity in our sampling of the number of Cepheids across the sky.
In the future, Cepheids will continue to be employed to investigate Galactic structure on large scales, as has been done in the past. For example, Cepheids have also been used to study the shape of the warp \citep{Skowron2019a,Skowron2019b,Chen:2019, Poggio2024arXiv} and its associated kinematics \citep{Poggio2018,Dehnen2023,HrannarJonsson:2024}.
To take full advantage of the large-scale coverage provided by the Cepheids, this incomplete and inhomogeneous dataset will need to be modelled taking into account both extinction and the selection function, that is, the expectation of the fraction of Cepheids in the dataset as a function of observables (e.g. $l,b,G,G$-$Rp$, and other quantities) to properly infer the intrinsic properties of the population \citep[][]{Cantat-Gaudin:2023,Cantat-Gaudin:2024,Khanna2024}.

The clearest and most prominent arm in our dataset, being immediately visible in both the $XY$ and $\phi^\prime-\ln R$ plots is the Sagittarius-Carina (Sgr-Car) arm. The spiral arm parameters derived for this arm account for the two tangent directions that have been traditionally attributed to the Sgr-Car arm. We identified the second most obvious spiral feature in the third quadrant with the Perseus arm because this spiral arm passes through the large Cassiopeia star-forming region that has long been identified as the nearest segment of the Perseus arm. However, consistent with earlier works, we find that this arm has a significantly larger pitch angle than derived from the masers in the second quadrant. 

Being intrinsically bright, the Cepheids allow us to map young stellar populations over a considerable extent of the Galaxy and yield an excellent complimentary dataset to the masers that provides us with information in the third and fourth quadrants, where to date only a few masers have measured astrometric parallaxes. In the first and second quadrants, the maser detection is less hindered by extinction effects than the Cepheids, which can only be reliably identified in the optical.  Based on these two populations and future surveys, we look forward to the large-scale structure of our Galaxy being more fully revealed as additional maser parallaxes become available from the Very Long Baseline Interferometry (VLBI) measurements in the southern hemisphere and as deep all-sky multi-epoch photometric surveys allow us to identify Cepheids that are currently hidden by interstellar extinction. 


\begin{acknowledgements}
We thank Alessandro Spagna and Robert Benjamin for useful discussions.

RD and EP are supported in part by the Italian Space Agency (ASI) through contract 2018-24-HH.0 and its addendum 2018-24-HH.1-2022 to the National Institute for Astrophysics (INAF).

SK \& RD acknowledge support from the European Union's Horizon 2020 research and innovation program under the GaiaUnlimited project (grant agreement No 101004110).

DMS acknowledges support from the European Union (ERC, LSP-MIST, 101040160). Views and opinions expressed are however those of the authors only and do not necessarily reflect those of the European Union or the European Research Council. Neither the European Union nor the granting authority can be held responsible for them.

SK acknowledges use of the INAF PLEIADI@IRA computing resources.

This work presents results from the European Space Agency (ESA) space mission \textit{Gaia}. \textit{Gaia} data are being processed by the \textit{Gaia} Data Processing and Analysis Consortium (DPAC). Funding for the DPAC is provided by national institutions, in particular the institutions participating in the \textit{Gaia} MultiLateral Agreement (MLA). The \textit{Gaia} mission website is https://www.cosmos.esa.int/gaia. The \textit{Gaia} archive website is https://archives.esac.esa.int/gaia.

This publication makes use of AllWISE data products derived from the Wide-field Infrared Survey Explorer, which is a joint project of the University of California, Los Angeles, and the Jet Propulsion Laboratory/California Institute of Technology, and NEOWISE, which is a project of the Jet Propulsion Laboratory/California Institute of Technology. WISE and NEOWISE are funded by the National Aeronautics and Space Administration.

This work has used the following software products:
Matplotlib \citep{Hunter:2007};
Astropy \citep{2018AJ....156..123A};
SciPy \citep{2020SciPy-NMeth};
and NumPy\citep{harris2020array}.
This project was developed in part at the Lorentz Center workshop "Mapping the Milky Way", held 6-10 February, 2023 in Leiden, Netherlands.

\end{acknowledgements}

   \bibliographystyle{aa} 
   \bibliography{mybib} 

\begin{thebibliography}{45}
\expandafter\ifx\csname natexlab\endcsname\relax\def\natexlab#1{#1}\fi

\bibitem[{{Anders} {et~al.}(2024){Anders}, {Padois}, {Vilanova Sar}, {Semczuk},
  {del Alc{\'a}zar}, \& {Figueras}}]{Anders:2024}
{Anders}, F., {Padois}, C., {Vilanova Sar}, M., {et~al.} 2024, arXiv e-prints,
  arXiv:2406.06228

\bibitem[{{Anderson} {et~al.}(2016){Anderson}, {Saio}, {Ekstr{\"o}m}, {Georgy},
  \& {Meynet}}]{Anderson2016}
{Anderson}, R.~I., {Saio}, H., {Ekstr{\"o}m}, S., {Georgy}, C., \& {Meynet}, G.
  2016, \aap, 591, A8

\bibitem[{{Astropy Collaboration} {et~al.}(2018){Astropy Collaboration},
  {Price-Whelan}, {Sip{\H o}cz}, {G{\"u}nther}, {Lim}, {Crawford}, {Conseil},
  {Shupe}, {Craig}, {Dencheva}, {Ginsburg}, {VanderPlas}, {Bradley},
  {P{\'e}rez-Su{\'a}rez}, {de Val-Borro}, {Aldcroft}, {Cruz}, {Robitaille},
  {Tollerud}, {Ardelean}, {Babej}, {Bach}, {Bachetti}, {Bakanov}, {Bamford},
  {Barentsen}, {Barmby}, {Baumbach}, {Berry}, {Biscani}, {Boquien}, {Bostroem},
  {Bouma}, {Brammer}, {Bray}, {Breytenbach}, {Buddelmeijer}, {Burke},
  {Calderone}, {Cano Rodr{\'{\i}}guez}, {Cara}, {Cardoso}, {Cheedella},
  {Copin}, {Corrales}, {Crichton}, {D'Avella}, {Deil}, {Depagne}, {Dietrich},
  {Donath}, {Droettboom}, {Earl}, {Erben}, {Fabbro}, {Ferreira}, {Finethy},
  {Fox}, {Garrison}, {Gibbons}, {Goldstein}, {Gommers}, {Greco}, {Greenfield},
  {Groener}, {Grollier}, {Hagen}, {Hirst}, {Homeier}, {Horton}, {Hosseinzadeh},
  {Hu}, {Hunkeler}, {Ivezi{\'c}}, {Jain}, {Jenness}, {Kanarek}, {Kendrew},
  {Kern}, {Kerzendorf}, {Khvalko}, {King}, {Kirkby}, {Kulkarni}, {Kumar},
  {Lee}, {Lenz}, {Littlefair}, {Ma}, {Macleod}, {Mastropietro}, {McCully},
  {Montagnac}, {Morris}, {Mueller}, {Mumford}, {Muna}, {Murphy}, {Nelson},
  {Nguyen}, {Ninan}, {N{\"o}the}, {Ogaz}, {Oh}, {Parejko}, {Parley}, {Pascual},
  {Patil}, {Patil}, {Plunkett}, {Prochaska}, {Rastogi}, {Reddy Janga},
  {Sabater}, {Sakurikar}, {Seifert}, {Sherbert}, {Sherwood-Taylor}, {Shih},
  {Sick}, {Silbiger}, {Singanamalla}, {Singer}, {Sladen}, {Sooley},
  {Sornarajah}, {Streicher}, {Teuben}, {Thomas}, {Tremblay}, {Turner},
  {Terr{\'o}n}, {van Kerkwijk}, {de la Vega}, {Watkins}, {Weaver}, {Whitmore},
  {Woillez}, {Zabalza}, \& {Astropy Contributors}}]{2018AJ....156..123A}
{Astropy Collaboration}, {Price-Whelan}, A., {Sip{\H o}cz}, B.~M., {et~al.}
  2018, {The Astropy Project: Building an Open-science Project and Status of
  the v2.0 Core Package}

\bibitem[{{Bobylev}(2022)}]{Bobylev2022}
{Bobylev}, V.~V. 2022, Astronomy Letters, 48, 126

\bibitem[{{Bono} {et~al.}(2024){Bono}, {Braga}, \& {Pietrinferni}}]{Bono2024}
{Bono}, G., {Braga}, V.~F., \& {Pietrinferni}, A. 2024, \aapr, 32, 4

\bibitem[{{Bronfman} {et~al.}(2000){Bronfman}, {Casassus}, {May}, \&
  {Nyman}}]{Bronfman:2000}
{Bronfman}, L., {Casassus}, S., {May}, J., \& {Nyman}, L.~{\r{A}}. 2000, \aap,
  358, 521

\bibitem[{Campello {et~al.}(2013)Campello, Moulavi, \& Sander}]{Campello2013}
Campello, R. J. G.~B., Moulavi, D., \& Sander, J. 2013, in Advances in
  Knowledge Discovery and Data Mining, ed. J.~Pei, V.~S. Tseng, L.~Cao,
  H.~Motoda, \& G.~Xu (Berlin, Heidelberg: Springer Berlin Heidelberg),
  160--172

\bibitem[{{Cantat-Gaudin} {et~al.}(2023){Cantat-Gaudin}, {Fouesneau}, {Rix},
  {Brown}, {Castro-Ginard}, {Kostrzewa-Rutkowska}, {Drimmel}, {Hogg}, {Casey},
  {Khanna}, {Oh}, {Price-Whelan}, {Belokurov}, {Saydjari}, \&
  {Green}}]{Cantat-Gaudin:2023}
{Cantat-Gaudin}, T., {Fouesneau}, M., {Rix}, H.-W., {et~al.} 2023, \aap, 669,
  A55

\bibitem[{{Cantat-Gaudin} {et~al.}(2024){Cantat-Gaudin}, {Fouesneau}, {Rix},
  {Brown}, {Drimmel}, {Castro-Ginard}, {Khanna}, {Belokurov}, \&
  {Casey}}]{Cantat-Gaudin:2024}
{Cantat-Gaudin}, T., {Fouesneau}, M., {Rix}, H.-W., {et~al.} 2024, \aap, 683,
  A128

\bibitem[{{Chen} {et~al.}(2019){Chen}, {Wang}, {Deng}, {de Grijs}, {Liu}, \&
  {Tian}}]{Chen:2019}
{Chen}, X., {Wang}, S., {Deng}, L., {et~al.} 2019, Nature Astronomy, 3, 320

\bibitem[{{Cutri} {et~al.}(2013){Cutri}, {Wright}, {Conrow}, {Fowler},
  {Eisenhardt}, {Grillmair}, {Kirkpatrick}, {Masci}, {McCallon}, {Wheelock},
  {Fajardo-Acosta}, {Yan}, {Benford}, {Harbut}, {Jarrett}, {Lake}, {Leisawitz},
  {Ressler}, {Stanford}, {Tsai}, {Liu}, {Helou}, {Mainzer}, {Gettings},
  {Gonzalez}, {Hoffman}, {Marsh}, {Padgett}, {Skrutskie}, {Beck}, {Papin}, \&
  {Wittman}}]{Cutri2013}
{Cutri}, R.~M., {Wright}, E.~L., {Conrow}, T., {et~al.} 2013, {Explanatory
  Supplement to the AllWISE Data Release Products}, Explanatory Supplement to
  the AllWISE Data Release Products, by R. M. Cutri et al.

\bibitem[{{Dehnen} {et~al.}(2023){Dehnen}, {Semczuk}, \&
  {Sch{\"o}nrich}}]{Dehnen2023}
{Dehnen}, W., {Semczuk}, M., \& {Sch{\"o}nrich}, R. 2023, \mnras, 523, 1556

\bibitem[{{Drimmel} {et~al.}(2023){Drimmel}, {Khanna}, {D'Onghia},
  {Tepper-Garc{\'\i}a}, {Bland-Hawthorn}, {Chemin}, {Ripepi},
  {Romero-G{\'o}mez}, {Ramos}, {Poggio}, {Andrae}, {Blomme}, {Cantat-Gaudin},
  {Castro-Ginard}, {Clementini}, {Figueras}, {Fouesneau}, {Fr{\'e}mat},
  {Lobel}, {Marshall}, \& {Muraveva}}]{Drimmel2023}
{Drimmel}, R., {Khanna}, S., {D'Onghia}, E., {et~al.} 2023, \aap, 670, A10

\bibitem[{{Gaia Collaboration} {et~al.}(2023){Gaia Collaboration}, {Drimmel},
  {Romero-G{\'o}mez}, {Chemin}, {Ramos}, {Poggio}, {Ripepi}, {Andrae},
  {Blomme}, {Cantat-Gaudin}, {Castro-Ginard}, {Clementini}, {Figueras},
  {Fouesneau}, {Fr{\'e}mat}, {Jardine}, {Khanna}, {Lobel}, {Marshall},
  {Muraveva}, {Brown}, {Vallenari}, {Prusti}, {de Bruijne}, {Arenou},
  {Babusiaux}, {Biermann}, {Creevey}, {Ducourant}, {Evans}, {Eyer}, {Guerra},
  {Hutton}, {Jordi}, {Klioner}, {Lammers}, {Lindegren}, {Luri}, {Mignard},
  {Panem}, {Pourbaix}, {Randich}, {Sartoretti}, {Soubiran}, {Tanga}, {Walton},
  {Bailer-Jones}, {Bastian}, {Jansen}, {Katz}, {Lattanzi}, {van Leeuwen},
  {Bakker}, {Cacciari}, {Casta{\~n}eda}, {De Angeli}, {Fabricius}, {Galluccio},
  {Guerrier}, {Heiter}, {Masana}, {Messineo}, {Mowlavi}, {Nicolas},
  {Nienartowicz}, {Pailler}, {Panuzzo}, {Riclet}, {Roux}, {Seabroke}, {Sordo},
  {Th{\'e}venin}, {Gracia-Abril}, {Portell}, {Teyssier}, {Altmann}, {Audard},
  {Bellas-Velidis}, {Benson}, {Berthier}, {Burgess}, {Busonero}, {Busso},
  {C{\'a}novas}, {Carry}, {Cellino}, {Cheek}, {Damerdji}, {Davidson}, {de
  Teodoro}, {Nu{\~n}ez Campos}, {Delchambre}, {Dell'Oro}, {Esquej},
  {Fern{\'a}ndez-Hern{\'a}ndez}, {Fraile}, {Garabato}, {Garc{\'\i}a-Lario},
  {Gosset}, {Haigron}, {Halbwachs}, {Hambly}, {Harrison}, {Hern{\'a}ndez},
  {Hestroffer}, {Hodgkin}, {Holl}, {Jan{\ss}en}, {Jevardat de Fombelle},
  {Jordan}, {Krone-Martins}, {Lanzafame}, {L{\"o}ffler}, {Marchal}, {Marrese},
  {Moitinho}, {Muinonen}, {Osborne}, {Pancino}, {Pauwels}, {Recio-Blanco},
  {Reyl{\'e}}, {Riello}, {Rimoldini}, {Roegiers}, {Rybizki}, {Sarro}, {Siopis},
  {Smith}, {Sozzetti}, {Utrilla}, {van Leeuwen}, {Abbas}, {{\'A}brah{\'a}m},
  {Abreu Aramburu}, {Aerts}, {Aguado}, {Ajaj}, {Aldea-Montero}, {Altavilla},
  {{\'A}lvarez}, {Alves}, {Anders}, {Anderson}, {Anglada Varela}, {Antoja},
  {Baines}, {Baker}, {Balaguer-N{\'u}{\~n}ez}, {Balbinot}, {Balog}, {Barache},
  {Barbato}, {Barros}, {Barstow}, {Bartolom{\'e}}, {Bassilana}, {Bauchet},
  {Becciani}, {Bellazzini}, {Berihuete}, {Bernet}, {Bertone}, {Bianchi},
  {Binnenfeld}, {Blanco-Cuaresma}, {Boch}, {Bombrun}, {Bossini}, {Bouquillon},
  {Bragaglia}, {Bramante}, {Breedt}, {Bressan}, {Brouillet}, {Brugaletta},
  {Bucciarelli}, {Burlacu}, {Butkevich}, {Buzzi}, {Caffau}, {Cancelliere},
  {Carballo}, {Carlucci}, {Carnerero}, {Carrasco}, {Casamiquela}, {Castellani},
  {Chaoul}, {Charlot}, {Chiaramida}, {Chiavassa}, {Chornay}, {Comoretto},
  {Contursi}, {Cooper}, {Cornez}, {Cowell}, {Crifo}, {Cropper}, {Crosta},
  {Crowley}, {Dafonte}, {Dapergolas}, {David}, {de Laverny}, {De Luise}, {De
  March}, {De Ridder}, {de Souza}, {de Torres}, {del Peloso}, {del Pozo},
  {Delbo}, {Delgado}, {Delisle}, {Demouchy}, {Dharmawardena}, {Di Matteo},
  {Diakite}, {Diener}, {Distefano}, {Dolding}, {Enke}, {Fabre}, {Fabrizio},
  {Faigler}, {Fedorets}, {Fernique}, {Fournier}, {Fouron}, {Fragkoudi}, {Gai},
  {Garcia-Gutierrez}, {Garcia-Reinaldos}, {Garc{\'\i}a-Torres}, {Garofalo},
  {Gavel}, {Gavras}, {Gerlach}, {Geyer}, {Giacobbe}, {Gilmore}, {Girona},
  {Giuffrida}, {Gomel}, {Gomez}, {Gonz{\'a}lez-N{\'u}{\~n}ez},
  {Gonz{\'a}lez-Santamar{\'\i}a}, {Gonz{\'a}lez-Vidal}, {Granvik}, {Guillout},
  {Guiraud}, {Guti{\'e}rrez-S{\'a}nchez}, {Guy}, {Hatzidimitriou}, {Hauser},
  {Haywood}, {Helmer}, {Helmi}, {Sarmiento}, {Hidalgo}, {H{\l}adczuk}, {Hobbs},
  {Holland}, {Huckle}, {Jasniewicz}, {Jean-Antoine Piccolo},
  {Jim{\'e}nez-Arranz}, {Juaristi Campillo}, {Julbe}, {Karbevska}, {Kervella},
  {Kordopatis}, {Korn}, {K{\'o}sp{\'a}l}, {Kostrzewa-Rutkowska},
  {Kruszy{\'n}ska}, {Kun}, {Laizeau}, {Lambert}, {Lanza}, {Lasne}, {Le
  Campion}, {Lebreton}, {Lebzelter}, {Leccia}, {Leclerc}, {Lecoeur-Taibi},
  {Liao}, {Licata}, {Lindstr{\o}m}, {Lister}, {Livanou}, {Lorca}, {Loup},
  {Madrero Pardo}, {Magdaleno Romeo}, {Managau}, {Mann}, {Manteiga},
  {Marchant}, {Marconi}, {Marcos}, {Marcos Santos}, {Mar{\'\i}n Pina},
  {Marinoni}, {Marocco}, {Martin Polo}, {Mart{\'\i}n-Fleitas}, {Marton},
  {Mary}, {Masip}, {Massari}, {Mastrobuono-Battisti}, {Mazeh}, {McMillan},
  {Messina}, {Michalik}, {Millar}, {Mints}, {Molina}, {Molinaro}, {Moln{\'a}r},
  {Monari}, {Mongui{\'o}}, {Montegriffo}, {Montero}, {Mor}, {Mora},
  {Morbidelli}, {Morel}, {Morris}, {Murphy}, {Musella}, {Nagy}, {Noval},
  {Oca{\~n}a}, {Ogden}, {Ordenovic}, {Osinde}, {Pagani}, {Pagano}, {Palaversa},
  {Palicio}, {Pallas-Quintela}, {Panahi}, {Payne-Wardenaar}, {Pe{\~n}alosa
  Esteller}, {Penttil{\"a}}, {Pichon}, {Piersimoni}, {Pineau}, {Plachy},
  {Plum}, {Pr{\v{s}}a}, {Pulone}, {Racero}, {Ragaini}, {Rainer}, {Raiteri},
  {Ramos-Lerate}, {Re Fiorentin}, {Regibo}, {Richards}, {Rios Diaz}, {Riva},
  {Rix}, {Rixon}, {Robichon}, {Robin}, {Robin}, {Roelens}, {Rogues},
  {Rohrbasser}, {Rowell}, {Royer}, {Ruz Mieres}, {Rybicki}, {Sadowski},
  {S{\'a}ez N{\'u}{\~n}ez}, {Sagrist{\`a} Sell{\'e}s}, {Sahlmann}, {Salguero},
  {Samaras}, {Sanchez Gimenez}, {Sanna}, {Santove{\~n}a}, {Sarasso},
  {Schultheis}, {Sciacca}, {Segol}, {Segovia}, {S{\'e}gransan}, {Semeux},
  {Shahaf}, {Siddiqui}, {Siebert}, {Siltala}, {Silvelo}, {Slezak}, {Slezak},
  {Smart}, {Snaith}, {Solano}, {Solitro}, {Souami}, {Souchay}, {Spagna},
  {Spina}, {Spoto}, {Steele}, {Steidelm{\"u}ller}, {Stephenson}, {S{\"u}veges},
  {Surdej}, {Szabados}, {Szegedi-Elek}, {Taris}, {Taylor}, {Teixeira},
  {Tolomei}, {Tonello}, {Torra}, {Torra}, {Torralba Elipe}, {Trabucchi},
  {Tsounis}, {Turon}, {Ulla}, {Unger}, {Vaillant}, {van Dillen}, {van Reeven},
  {Vanel}, {Vecchiato}, {Viala}, {Vicente}, {Voutsinas}, {Weiler}, {Wevers},
  {Wyrzykowski}, {Yoldas}, {Yvard}, {Zhao}, {Zorec}, {Zucker}, \&
  {Zwitter}}]{Drimmel23_gaia}
{Gaia Collaboration}, {Drimmel}, R., {Romero-G{\'o}mez}, M., {et~al.} 2023,
  \aap, 674, A37

\bibitem[{{GRAVITY Collaboration} {et~al.}(2022){GRAVITY Collaboration},
  {Abuter}, {Aimar}, {Amorim}, {Ball}, {Baub{\"o}ck}, {Berger}, {Bonnet},
  {Bourdarot}, {Brandner}, {Cardoso}, {Cl{\'e}net}, {Dallilar}, {Davies}, {de
  Zeeuw}, {Dexter}, {Drescher}, {Eisenhauer}, {F{\"o}rster Schreiber},
  {Foschi}, {Garcia}, {Gao}, {Gendron}, {Genzel}, {Gillessen}, {Habibi},
  {Haubois}, {Hei{\ss}el}, {Henning}, {Hippler}, {Horrobin}, {Jochum}, {Jocou},
  {Kaufer}, {Kervella}, {Lacour}, {Lapeyr{\`e}re}, {Le Bouquin}, {L{\'e}na},
  {Lutz}, {Ott}, {Paumard}, {Perraut}, {Perrin}, {Pfuhl}, {Rabien},
  {Shangguan}, {Shimizu}, {Scheithauer}, {Stadler}, {Stephens}, {Straub},
  {Straubmeier}, {Sturm}, {Tacconi}, {Tristram}, {Vincent}, {von Fellenberg},
  {Widmann}, {Wieprecht}, {Wiezorrek}, {Woillez}, {Yazici}, \&
  {Young}}]{GravityCollaboration:2022}
{GRAVITY Collaboration}, {Abuter}, R., {Aimar}, N., {et~al.} 2022, \aap, 657,
  L12

\bibitem[{{Griv} {et~al.}(2017){Griv}, {Jiang}, \& {Hou}}]{Griv2017}
{Griv}, E., {Jiang}, I.-G., \& {Hou}, L.-G. 2017, \apj, 844, 118

\bibitem[{{Haffner} {et~al.}(1999){Haffner}, {Reynolds}, \&
  {Tufte}}]{Haffner1999}
{Haffner}, L.~M., {Reynolds}, R.~J., \& {Tufte}, S.~L. 1999, \apj, 523, 223

\bibitem[{Harris {et~al.}(2020)Harris, Millman, van~der Walt, Gommers,
  Virtanen, Cournapeau, Wieser, Taylor, Berg, Smith, Kern, Picus, Hoyer, van
  Kerkwijk, Brett, Haldane, del R{\'{i}}o, Wiebe, Peterson,
  G{\'{e}}rard-Marchant, Sheppard, Reddy, Weckesser, Abbasi, Gohlke, \&
  Oliphant}]{harris2020array}
Harris, C.~R., Millman, K.~J., van~der Walt, S.~J., {et~al.} 2020, Array
  programming with {NumPy}

\bibitem[{Hunter(2007)}]{Hunter:2007}
Hunter, J.~D. 2007, Matplotlib: A 2D graphics environment

\bibitem[{{J{\'o}nsson} \& {McMillan}(2024)}]{HrannarJonsson:2024}
{J{\'o}nsson}, V.~H. \& {McMillan}, P.~J. 2024, \aap, 688, A38

\bibitem[{{Khanna} {et~al.}(2024){Khanna}, {Yu}, {Drimmel}, {Poggio},
  {Cantat-Gaudin}, {Castro-Ginard}, {Kurbatov}, {Belokurov}, {Brown},
  {Fouesneau}, {Casey}, \& {Rix}}]{Khanna2024}
{Khanna}, S., {Yu}, J., {Drimmel}, R., {et~al.} 2024, arXiv e-prints,
  arXiv:2410.22036

\bibitem[{{Kuhn} {et~al.}(2021){Kuhn}, {Benjamin}, {Zucker}, {Krone-Martins},
  {de Souza}, {Castro-Ginard}, {Ishida}, {Povich}, \&
  {Hillenbrand}}]{Kuhn:2021}
{Kuhn}, M.~A., {Benjamin}, R.~A., {Zucker}, C., {et~al.} 2021, \aap, 651, L10

\bibitem[{{Lemasle} {et~al.}(2022){Lemasle}, {Lala}, {Kovtyukh}, {Hanke},
  {Prudil}, {Bono}, {Braga}, {da Silva}, {Fabrizio}, {Fiorentino},
  {Fran{\c{c}}ois}, {Grebel}, \& {Kniazev}}]{Lemasle2022}
{Lemasle}, B., {Lala}, H.~N., {Kovtyukh}, V., {et~al.} 2022, \aap, 668, A40

\bibitem[{{Levine} {et~al.}(2006){Levine}, {Blitz}, \& {Heiles}}]{Levine06}
{Levine}, E.~S., {Blitz}, L., \& {Heiles}, C. 2006, Science, 312, 1773

\bibitem[{{Minniti} {et~al.}(2021){Minniti}, {Zoccali}, {Rojas-Arriagada},
  {Minniti}, {Sbordone}, {Contreras Ramos}, {Braga}, {Catelan}, {Duffau},
  {Gieren}, {Marconi}, \& {Valcarce}}]{Minniti2021}
{Minniti}, J.~H., {Zoccali}, M., {Rojas-Arriagada}, A., {et~al.} 2021, \aap,
  654, A138

\bibitem[{{Morgan} {et~al.}(1953){Morgan}, {Whitford}, \& {Code}}]{Morgan1953}
{Morgan}, W.~W., {Whitford}, A.~E., \& {Code}, A.~D. 1953, \apj, 118, 318

\bibitem[{Pedregosa {et~al.}(2011)Pedregosa, Varoquaux, Gramfort, Michel,
  Thirion, Grisel, Blondel, Prettenhofer, Weiss, Dubourg, Vanderplas, Passos,
  Cournapeau, Brucher, Perrot, \& Duchesnay}]{scikit-learn}
Pedregosa, F., Varoquaux, G., Gramfort, A., {et~al.} 2011, Journal of Machine
  Learning Research, 12, 2825

\bibitem[{{Pietrukowicz} {et~al.}(2021){Pietrukowicz}, {Soszy{\'n}ski}, \&
  {Udalski}}]{Pietrukowicz:2021}
{Pietrukowicz}, P., {Soszy{\'n}ski}, I., \& {Udalski}, A. 2021, \actaa, 71, 205

\bibitem[{{Poggio} {et~al.}(2021){Poggio}, {Drimmel}, {Cantat-Gaudin}, {Ramos},
  {Ripepi}, {Zari}, {Andrae}, {Blomme}, {Chemin}, {Clementini}, {Figueras},
  {Fouesneau}, {Fr{\'e}mat}, {Lobel}, {Marshall}, {Muraveva}, \&
  {Romero-G{\'o}mez}}]{Poggio2021}
{Poggio}, E., {Drimmel}, R., {Cantat-Gaudin}, T., {et~al.} 2021, \aap, 651,
  A104

\bibitem[{{Poggio} {et~al.}(2018){Poggio}, {Drimmel}, {Lattanzi}, {Smart},
  {Spagna}, {Andrae}, {Bailer-Jones}, {Fouesneau}, {Antoja}, {Babusiaux},
  {Evans}, {Figueras}, {Katz}, {Reyl{\'e}}, {Robin}, {Romero-G{\'o}mez}, \&
  {Seabroke}}]{Poggio2018}
{Poggio}, E., {Drimmel}, R., {Lattanzi}, M.~G., {et~al.} 2018, \mnras, 481, L21

\bibitem[{{Poggio} {et~al.}(2024){Poggio}, {Khanna}, {Drimmel}, {Zari},
  {D'Onghia}, {Lattanzi}, {Palicio}, {Recio-Blanco}, \&
  {Thulasidharan}}]{Poggio2024arXiv}
{Poggio}, E., {Khanna}, S., {Drimmel}, R., {et~al.} 2024, arXiv e-prints,
  arXiv:2407.18659

\bibitem[{{Reid} {et~al.}(2019){Reid}, {Menten}, {Brunthaler}, {Zheng}, {Dame},
  {Xu}, {Li}, {Sakai}, {Wu}, {Immer}, {Zhang}, {Sanna}, {Moscadelli}, {Rygl},
  {Bartkiewicz}, {Hu}, {Quiroga-Nu{\~n}ez}, \& {van Langevelde}}]{Reid2019}
{Reid}, M.~J., {Menten}, K.~M., {Brunthaler}, A., {et~al.} 2019, \apj, 885, 131

\bibitem[{{Schlafly} {et~al.}(2019){Schlafly}, {Meisner}, \&
  {Green}}]{Schlafly2019}
{Schlafly}, E.~F., {Meisner}, A.~M., \& {Green}, G.~M. 2019, \apjs, 240, 30

\bibitem[{{Skowron} {et~al.}(2025){Skowron}, {Drimmel}, {Khanna}, {Spagna},
  {Poggio}, \& {Ramos}}]{Skowron2024}
{Skowron}, D.~M., {Drimmel}, R., {Khanna}, S., {et~al.} 2025, \apjs, 278, 57

\bibitem[{{Skowron} {et~al.}(2019{\natexlab{a}}){Skowron}, {Skowron},
  {Mr{\'o}z}, {Udalski}, {Pietrukowicz}, {Soszy{\'n}ski}, {Szyma{\'n}ski},
  {Poleski}, {Koz{\l}owski}, {Ulaczyk}, {Rybicki}, \& {Iwanek}}]{Skowron2019a}
{Skowron}, D.~M., {Skowron}, J., {Mr{\'o}z}, P., {et~al.} 2019{\natexlab{a}},
  Science, 365, 478

\bibitem[{{Skowron} {et~al.}(2019{\natexlab{b}}){Skowron}, {Skowron},
  {Mr{\'o}z}, {Udalski}, {Pietrukowicz}, {Soszy{\'n}ski}, {Szyma{\'n}ski},
  {Poleski}, {Koz{\l}owski}, {Ulaczyk}, {Rybicki}, {Iwanek}, {. Wrona}, \&
  {Gromadzki}}]{Skowron2019b}
{Skowron}, D.~M., {Skowron}, J., {Mr{\'o}z}, P., {et~al.} 2019{\natexlab{b}},
  \actaa, 69, 305

\bibitem[{{van de Hulst} {et~al.}(1954){van de Hulst}, {Muller}, \&
  {Oort}}]{vandHulst1954}
{van de Hulst}, H.~C., {Muller}, C.~A., \& {Oort}, J.~H. 1954, \bain, 12, 117

\bibitem[{van~der Maaten \& Hinton(2008)}]{vandermaaten2008}
van~der Maaten, L. \& Hinton, G. 2008, Journal of Machine Learning Research, 9,
  2579

\bibitem[{{VERA Collaboration} {et~al.}(2020){VERA Collaboration}, {Hirota},
  {Nagayama}, {Honma}, {Adachi}, {Burns}, {Chibueze}, {Choi}, {Hachisuka},
  {Hada}, {Hagiwara}, {Hamada}, {Handa}, {Hashimoto}, {Hirano}, {Hirata},
  {Ichikawa}, {Imai}, {Inenaga}, {Ishikawa}, {Jike}, {Kameya}, {Kaseda}, {Kim},
  {Kim}, {Kim}, {Kobayashi}, {Kono}, {Kurayama}, {Matsuno}, {Morita}, {Motogi},
  {Murase}, {Nakagawa}, {Nakanishi}, {Niinuma}, {Nishi}, {Oh}, {Omodaka},
  {Oyadomari}, {Oyama}, {Sakai}, {Sakai}, {Sawada-Satoh}, {Shibata},
  {Shizugami}, {Sudo}, {Sugiyama}, {Sunada}, {Suzuki}, {Takahashi}, {Tamura},
  {Tazaki}, {Ueno}, {Uno}, {Urago}, {Wada}, {Wu}, {Yamashita}, {Yamashita},
  {Yamauchi}, \& {Yuda}}]{VERA2020}
{VERA Collaboration}, {Hirota}, T., {Nagayama}, T., {et~al.} 2020, \pasj, 72,
  50

\bibitem[{Virtanen {et~al.}(2020)Virtanen, Gommers, Oliphant, Haberland, Reddy,
  Cournapeau, Burovski, Peterson, Weckesser, Bright, {van der Walt}, Brett,
  Wilson, Millman, Mayorov, Nelson, Jones, Kern, Larson, Carey, Polat, Feng,
  Moore, {VanderPlas}, Laxalde, Perktold, Cimrman, Henriksen, Quintero, Harris,
  Archibald, Ribeiro, Pedregosa, {van Mulbregt}, \& {SciPy 1.0
  Contributors}}]{2020SciPy-NMeth}
Virtanen, P., Gommers, R., Oliphant, T.~E., {et~al.} 2020, {{SciPy} 1.0:
  Fundamental Algorithms for Scientific Computing in Python}

\bibitem[{{Wright} {et~al.}(2010){Wright}, {Eisenhardt}, {Mainzer}, {Ressler},
  {Cutri}, {Jarrett}, {Kirkpatrick}, {Padgett}, {McMillan}, {Skrutskie},
  {Stanford}, {Cohen}, {Walker}, {Mather}, {Leisawitz}, {Gautier}, {McLean},
  {Benford}, {Lonsdale}, {Blain}, {Mendez}, {Irace}, {Duval}, {Liu}, {Royer},
  {Heinrichsen}, {Howard}, {Shannon}, {Kendall}, {Walsh}, {Larsen}, {Cardon},
  {Schick}, {Schwalm}, {Abid}, {Fabinsky}, {Naes}, \& {Tsai}}]{Wright2010}
{Wright}, E.~L., {Eisenhardt}, P. R.~M., {Mainzer}, A.~K., {et~al.} 2010, \aj,
  140, 1868

\bibitem[{{Xu} {et~al.}(2018){Xu}, {Bian}, {Reid}, {Li}, {Zhang}, {Yan},
  {Dame}, {Menten}, {He}, {Liao}, \& {Tang}}]{Xu:2018}
{Xu}, Y., {Bian}, S.~B., {Reid}, M.~J., {et~al.} 2018, \aap, 616, L15

\bibitem[{{Xu} {et~al.}(2023){Xu}, {Hao}, {Liu}, {Lin}, {Bian}, {Hou}, {Li}, \&
  {Li}}]{Xu2023}
{Xu}, Y., {Hao}, C.~J., {Liu}, D.~J., {et~al.} 2023, \apj, 947, 54

\bibitem[{{Xu} {et~al.}(2016){Xu}, {Reid}, {Dame}, {Menten}, {Sakai}, {Li},
  {Brunthaler}, {Moscadelli}, {Zhang}, \& {Zheng}}]{Xu2016}
{Xu}, Y., {Reid}, M., {Dame}, T., {et~al.} 2016, Science Advances, 2, e1600878

\bibitem[{{Zari} {et~al.}(2021){Zari}, {Rix}, {Frankel}, {Xiang}, {Poggio},
  {Drimmel}, \& {Tkachenko}}]{Zari2021}
{Zari}, E., {Rix}, H.~W., {Frankel}, N., {et~al.} 2021, \aap, 650, A112

\end{thebibliography}

\end{document}